%% file: ms.tex
\documentclass[12pt,twoside,english]{article}
\usepackage{a4wide}
\pdfoutput=1

\makeatletter

\input{settings/settings_tools}
\input{settings/mycommands}

\newcommand{\lyxaddress}[1]{
\par {\raggedright #1
\vspace{1.4em}
\noindent\par}
}

\numberwithin{equation}{section}

\makeatother

\graphicspath{ {./pics/} }
\newcommand{\revStart}{\color{black}}
\newcommand{\revEnd}{\color{black}}

\def\pathToBibFile{references}

\begin{document}

\title{\vspace{-2cm}Kirchhoff-Love shell theory based on tangential differential calculus}

\author{D. Sch{\"o}llhammer, T.P. Fries}
\maketitle

\lyxaddress{\begin{center}
Institute of Structural Analysis\\
Graz University of Technology\\
Lessingstr. 25/II, 8010 Graz, Austria\\
\texttt{www.ifb.tugraz.at}\\
\texttt{schoellhammer@tugraz.at}
\end{center}}

\input{Content/Abstract}
\input{Content/Introduction}
\input{Content/Preliminaries}
\input{Content/Methodology}
\input{Content/Implementation}
\input{Content/NumRes}
\input{Content/Conclusion}

\newpage
\appendix
\input{Content/AppendixI}

\clearpage
\newpage

\bibliographystyle{schanz}
\addcontentsline{toc}{section}{\refname}
\bibliography{\pathToBibFile}
 
\end{document}

%% file: settings/settings_tools.tex
\usepackage[utf8]{inputenc}            
\usepackage[T1]{fontenc} 			
\usepackage{graphicx}
\usepackage{fancyhdr}
\usepackage{amsmath, amssymb, amsfonts, bm}
\usepackage{caption}
\usepackage[labelformat=simple]{subfig}
\usepackage{setspace}
\usepackage{verbatim} 
\usepackage[english]{babel}
\usepackage{pdfsync}
\usepackage{graphicx}
\usepackage{caption}
\usepackage{booktabs}
\usepackage{hyperref}
\usepackage{float}

\usepackage{siunitx}
\usepackage{xfrac}
\usepackage{pdfpages}

\usepackage{color}
\usepackage{listings}

\includepdfset{turn=true,pagecommand={}}

\hypersetup{
	colorlinks = false,
	linktocpage = true,
	allbordercolors = {0.8 0.8 0.8}
}

\addto\captionsenglish{}
\addto\captionsenglish{}
\addto\extrasenglish{}
\addto\extrasenglish{}
\addto\extrasenglish{}
\addto\extrasenglish{}
\addto\extrasenglish{}
\addto\extrasenglish{}
\addto\extrasenglish{}
\addto\extrasenglish{}

\usepackage{etex}
\usepackage{tikz}
\usepackage{pgfplots}
\usepackage{tikz-3dplot}

\pgfplotsset{compat=newest}

\ifdefined\bUseTikzExternalize
	\usetikzlibrary{external}
	\tikzsetexternalprefix{tmp/}
\else
	\usetikzlibrary{external}
	\tikzexternaldisable
\fi

\usetikzlibrary{calc,intersections}
\usetikzlibrary{shapes.geometric,shapes.arrows,decorations.pathmorphing}
\usetikzlibrary{matrix,chains,scopes,positioning,arrows,fit}
\usetikzlibrary{spy}
\usetikzlibrary{fadings}
\usetikzlibrary{shadings}
\usetikzlibrary{backgrounds}

\usepgfplotslibrary{groupplots}
\usepgfplotslibrary{patchplots}

\def\pgfplotfontsizetitle{\small}
\def\pgfplotfontsize{\small}

\pgfplotsset{
  mystyle/.style ={%
    grid = major,
    every tick label/.append style={font=\pgfplotfontsize},
    every axis label/.append style={font=\pgfplotfontsize},
    legend style={font=\scriptsize},
    label style={font=\pgfplotfontsize},
    title style={font=\pgfplotfontsizetitle},
    /pgf/number format/set thousands separator = {}, 
  }
}

\usepackage[outline]{contour}
\usepackage{standalone} 



%% file: settings/mycommands.tex
\newcommand{\vek}[1]{\mathchoice{\displaystyle\boldsymbol#1}
	{\textstyle\boldsymbol#1}{\scriptstyle\boldsymbol#1}
	{\scriptscriptstyle\boldsymbol#1}}
\newcommand{\mat}[1]{\mathchoice{\displaystyle\mathbf#1}
	{\textstyle\mathbf#1}{\scriptstyle\mathbf#1}
	{\scriptscriptstyle\mathbf#1}}

\newcommand{\ten}[1]{ \ensuremath{\bm #1} }
\renewcommand{\d}{ \ensuremath{\mathrm{d} }}
\newcommand{\p}{ \ensuremath{\partial} }
\renewcommand{\t}[1]{\text{#1}}
\newcommand{\T}{ \ensuremath{\intercal}}

\newcommand{\divG}[1]{\ensuremath{\text{div}_{\Gamma} #1}}
\newcommand{\gradR}[1]{\ensuremath{\nabla_{\vek{r}} #1}}
\newcommand{\gradG}[1]{\ensuremath{\nabla_\Gamma #1}}
\newcommand{\gradGD}[1]{\ensuremath{\nabla_\Gamma^\text{dir} #1}}
\newcommand{\gradGC}[1]{\ensuremath{\nabla_\Gamma^\text{cov} #1}}
\newcommand{\nG}{\ensuremath{\vek{n}_\Gamma}}
\newcommand{\nCo}{\ensuremath{\vek{n}_{\p\Gamma}}}
\newcommand{\tB}{\ensuremath{\vek{t}_{\p\Gamma}}}

\makeatletter
\newcommand*{\overlaynumber}{\number\beamer@slideinframe}
\makeatother

\newcommand{\tikzfigforpaper}[3]{
			\def\tkzscale{#2}
			\def\dirtikz{tikz} 
			\def\dirdata{data} 
			\centering
			\ifdefined\bUseTikzExternalize
			\tikzsetnextfilename{#3}  
			\else
				\includegraphics{#3}
			\fi			
	}

\tikzset{
	>=stealth',
	axis/.style={<->},
	important line/.style={thick},
	connection/.style={thick, dotted},
	dot/.style = {
		draw,
		fill = white,
		circle,
		solid,
		thin,
		inner sep = 0pt,
		minimum size = 3pt
	},
	defP/.style = {
		inner sep = 0pt
	}	
}
\pgfarrowsdeclaredouble{doubled}{doubled}{stealth'}{stealth'}

%% file: Content/Abstract.tex
\begin{abstract}
\revStart
The Kirchhoff-Love shell theory is recasted in the frame of the tangential differential calculus (TDC) where differential operators on surfaces are formulated based on global, three-dimensional coordinates. As a consequence, there is no need for a parametrization of the shell geometry implying curvilinear surface coordinates as used in the classical shell theory. Therefore, the proposed TDC-based formulation also applies to shell geometries which are zero-isosurfaces as in the level-set method where no parametrization is available in general. For the discretization, the TDC-based formulation may be used based on surface meshes implying element-wise parametrizations. Then, the results are equivalent to those obtained based on the classical theory. However, it may also be used in recent finite element approaches as the TraceFEM and CutFEM where shape functions are generated on a background mesh without any need for a parametrization. Numerical results presented herein are achieved with isogeometric analysis for classical and new benchmark tests. Higher-order convergence rates in the residual errors are achieved when the physical fields are sufficiently smooth. \par\medskip
\revEnd


\end{abstract} 
\newpage\tableofcontents\newpage

%% file: Content/Introduction.tex
\section{Introduction}
\label{sec:intro}

\revStart
The mechanical modeling of shells leads to partial differential equations (PDEs) on manifolds where the manifolds are curved surfaces in the three-dimensional space. An overview in classical shell theory is given, e.g., in \cite{Bischoff_2004a,Cirak_2000a,Kiendl_2009a,Simo_1989a,Simo_1989b} or in the textbooks \cite{Basar_1985a,Blaauwendraad_2014a,Radwanska_2017a,Wempner_2002a}. When modeling physical phenomena on curved surfaces, definitions for geometric quantities (normal vectors, curvatures, etc.) and differential surface operators (gradients, divergence, etc.) are key ingredients. These quantities may be either defined based on two-dimensional, curvilinear \emph{local} coordinates living on the manifold or on \emph{global} coordinates of the surrounding, three-dimensional space.\par
 
\begin{figure}[htb]
	\centering
	\subfloat[parametric via map $\vek{x}(\vek{r})$]{\includegraphics[width=.22\textwidth]{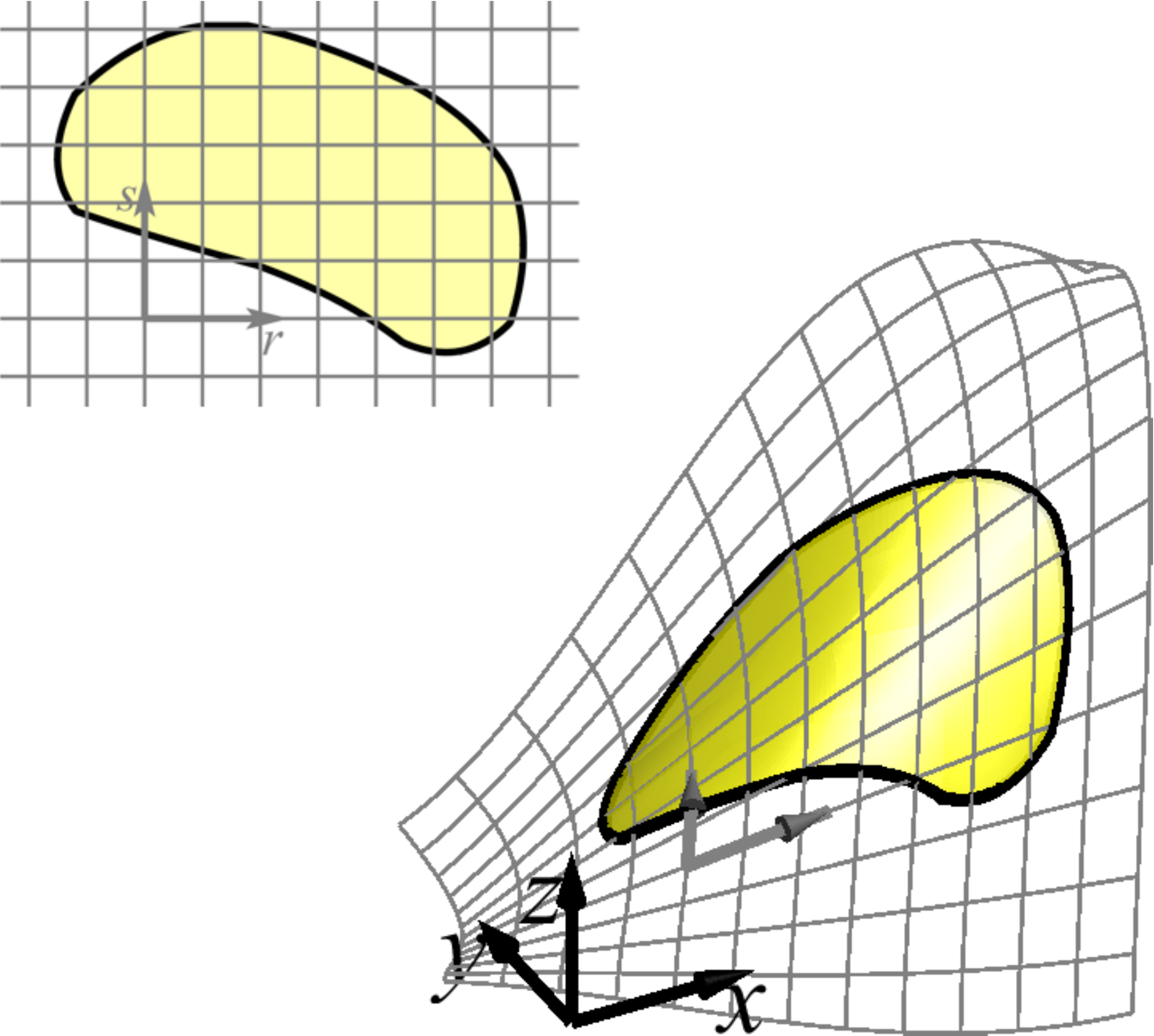}\label{fig:introA}} \hfill
	\subfloat[\revStart Shell-BVP on zero-isosurface \revEnd]{\includegraphics[width=.22\textwidth]{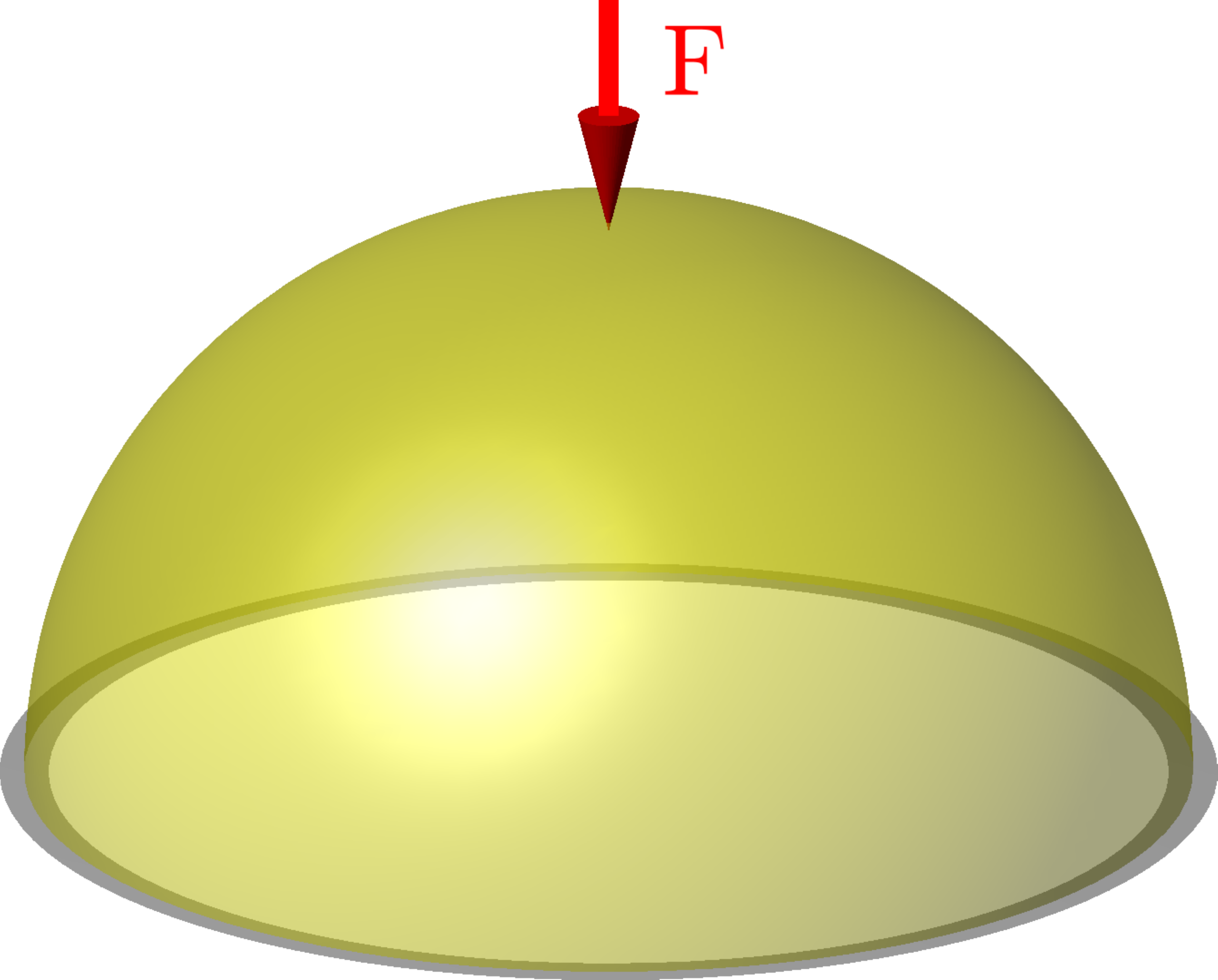}\label{fig:introB}} \hfill
	\subfloat[maps implied by FEM]{\includegraphics[width=.22\textwidth]{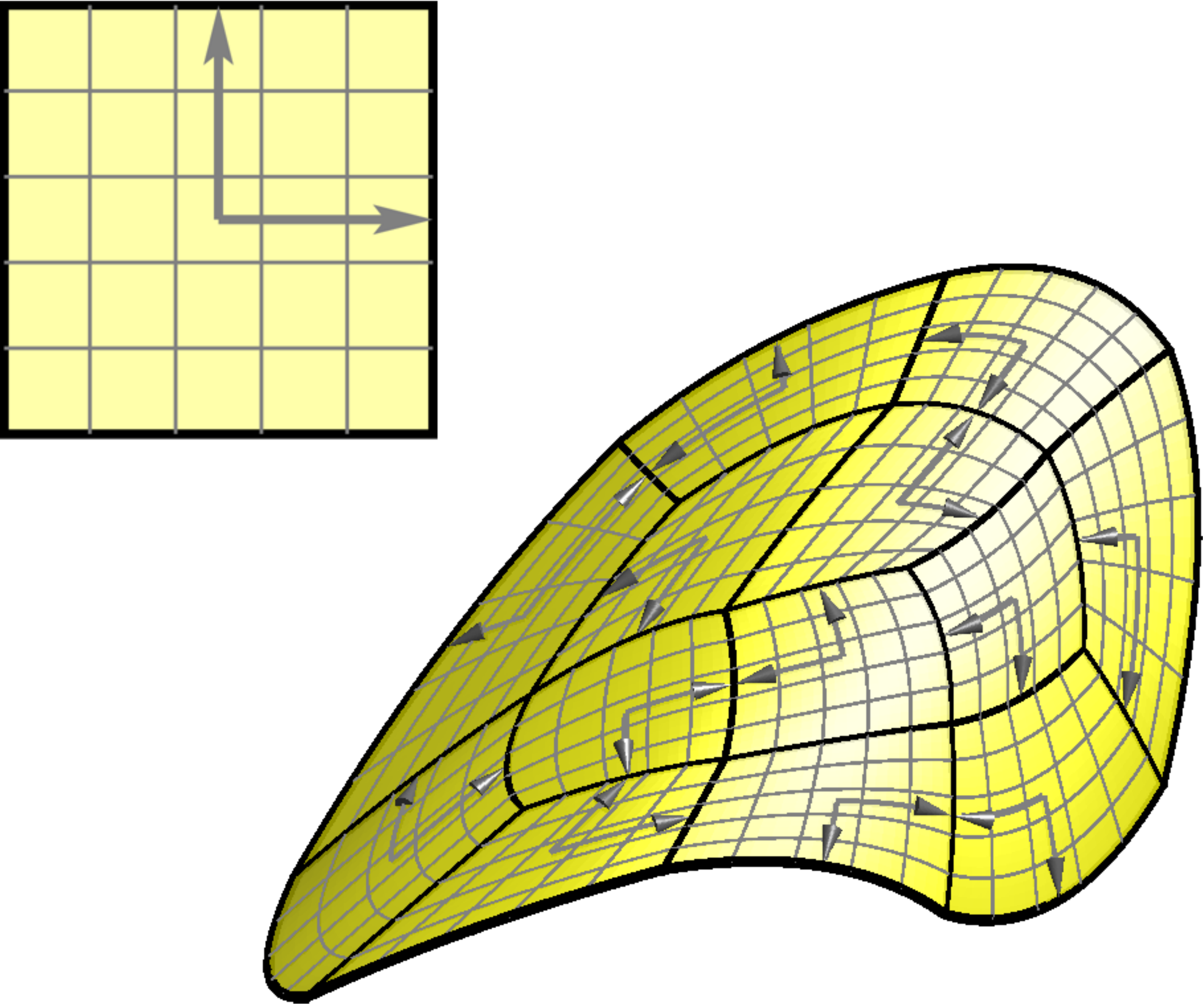}\label{fig:introC}} \hfill
	\subfloat[TraceFEM, CutFEM]{\includegraphics[width=.22\textwidth]{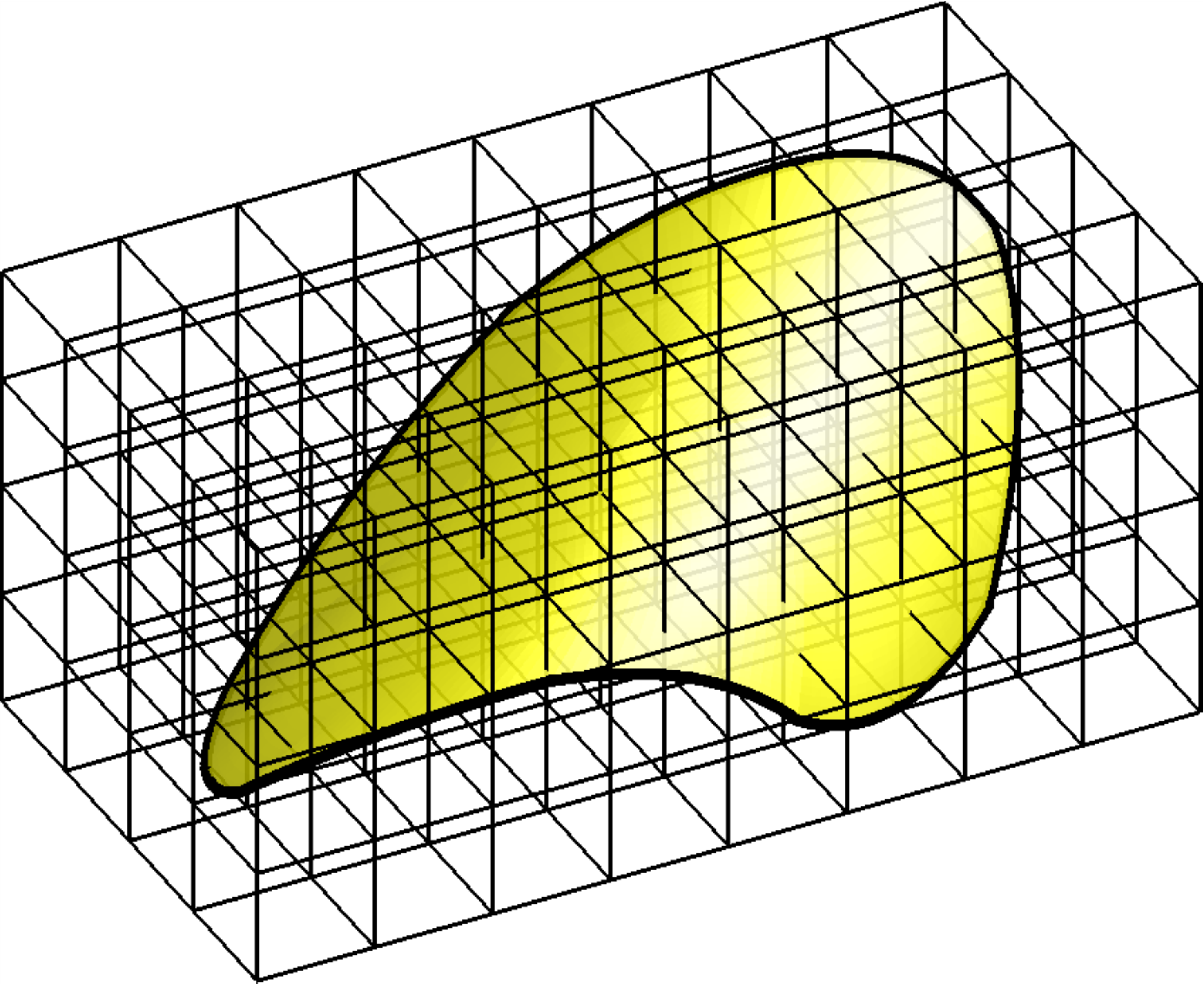}\label{fig:introD}}
	\caption{(a) In classical shell mechanics, the middle surface is defined by a parametrization, i.e., a map $\vek{x}(\vek{r})$. \revStart (b) The cupola is given by the zero-isosurface of $\phi(\vek{x})$ and the mechanical response to the force $F$ is sought. \revEnd (c) The surface mesh implies element-wise, approximate parametrizations even if the initial geometry is defined by level-sets. (d) For implicitly defined shells in the context of TraceFEM and CutFEM, no parametrization is needed at all.}
	\label{fig:intro}
\end{figure}

In the first case, the curved surface is parametrized by two parameters, i.e., there is a given map from the two-dimensional parameter space to the three-dimensional physical space, see \autoref{fig:introA}. For the definition of geometrical quantities and surface operators, co- and contra-variant base vectors and Christoffel-symbols naturally occur. It is important to note that a parametrization of a surface is not unique, hence, there are infinitely many maps which result in the same curved surface. Obviously, the physical modeling must be independent of a concrete parametrization, which suggests the existence of a parametrization-free formulation.\par

In the second case, the geometric quantities and surface operators are based on global coordinates as done in the tangential differential calculus (TDC) \cite{Delfour_2011a,Gurtin_1975a,Hansbo_2015a}. Then, a model may also be defined even if a parametrization of a curved surface does not exist, for example, when it is a zero-isosurface of a scalar function in three dimensions following the level-set method \cite{Fries_2017a,Fries_2017b,Osher_2003a,Sethian_1999b}. When the physical modeling is based on the TDC, i.e., on global coordinates, it is applicable to surfaces which are parametrized or not. In this sense, the TDC-based approach is more general than approaches based on local coordinates. Models based on the TDC are found in various applications, see \cite{Demlow_2009a,Dziuk_1988a,Dziuk_2013a,Fries_2017b} for scalar problems such as heat flow and \cite{Fries_2018b,Jankuhn_2017a} for flow problems on manifolds. In the context of structure mechanics, this approach is used in \cite{Hansbo_2014b} for curved beams, in \cite{Hansbo_2014a,Hansbo_2015a,Gurtin_1975a} for membranes, and in \cite{Hansbo_2017a} for flat shells embedded in $\mathbb{R}^3$.\par

Herein, we apply the TDC for the reformulation of the classical Kirchhoff-Love shell theory which is typically formulated based on a given parametrization. Based on the TDC, it is possible to also formulate the boundary value problem (BVP) for shell geometries where no parametrization is given as for the example in \autoref{fig:introB}: The cupola with radius r is given by the zero-isosurface of $\phi(\vek{x}) = \Vert\vek{x}\Vert-r$ with $\vek{x} \in [-r,r]^2 \times [0,r]$ and the mechanical response to the force $F$ is sought. As mentioned before, the TDC-based formulation is also valid when a parametrization is available; it is then equivalent to the classical formulation based on local coordinates. \par

Other attempts to parametrization-free formulations of the Kirchhoff-Love shell theory are found, e.g., in \cite{Delfour_1994a,Delfour_1995a,Delfour_1996a,Delfour_1997a} with a mathematical focus and in \cite{Lebiedzik_2007a,Opstal_2015a,Yao_2009a} from an engineering perspective, however, only with focus on displacements. Herein, the Kirchhoff-Love shell theory is recasted in the frame of the TDC including all relevant mechanical aspects. For the first time, the parametrization-free strong form of the Kirchhoff-Love shell is given and taken as the starting point to derive the weak form. Then, boundary terms for the relevant boundary conditions of Kirchhoff-Love shell theory are naturally achieved. Furthermore, mechanical quantities such as moments, normal and shear forces are defined based on global coordinates and it is shown how (parametrization-)invariant quantities such as principal moments are computed. Finally, the strong form of Kirchhoff-Love shells is also found highly useful to define residual errors in the numerical results. Of course, evaluating this error in the strong form requires up to forth-order derivatives on the surface, which is implementationally quite some effort. The advantage, however, is that one may then confirm higher-order convergence rates in the corresponding error norm for suitable shell test cases. This is, otherwise, very difficult as exact solutions for shells are hardly available and classical benchmark tests typically give only selected scalar quantities, often with moderate accuracy.\par

For the numerical solution of shells, i.e., the approximation of the shell BVP based on numerical methods, we distinguish two fundamentally different approaches. The first is a classical finite element analysis based on a surface mesh, labelled Surface FEM herein \cite{Demlow_2009a,Dziuk_2013a,Fries_2018b,Fries_2017b}. Once a surface mesh is generated, it implies element-wise parametrizations for the shell geometry, see \autoref{fig:introC}, no matter whether the underlying (analytic) geometry was parametrized or implied by level sets. In this case, classical shell theory based on parametrizations is suitable at least for the discretized geometry. The proposed TDC-based formulation is suitable as well which shall be seen in the numerical results. The other numerical approach is to use a three-dimensional background mesh into which the curved shell surface is embedded, cf.~\autoref{fig:introD}. Then, the shape functions of the (three-dimensional) background elements are only evaluated on the shell surface and no parametrization (and surface mesh) is needed to furnish basis functions for the approximation. For these methods, e.g., labelled CutFEM \cite{Burman_2015a,Burman_2018a,Cenanovic_2016a,Elfverson_2018a} or TraceFEM \cite{Grande_2016a,Olshanskii_2017a,Olshanskii_2017b,Reusken_2014a}, applied to the case of shell mechanics, it is no longer possible to rely on classical parametrization-based formulations of the shell mechanics, however, the proposed TDC-based formulation is still applicable. \par

For the numerical results presented herein, the continuous weak form of the BVP is discretized with the Surface FEM \cite{Demlow_2009a,Dziuk_2013a,Fries_2018b,Fries_2017b} using NURBS as trial and test functions as proposed by Hughes et al.~\cite{Cottrell_2009a,Hughes_2005a} due to the continuity requirements of Kirchhoff-Love shells. The boundary conditions are weakly enforced via Lagrange multiplies \cite{Zienkiewicz_2013a}. The situation is similar to \cite{Belytschko_1985a,Kiendl_2009a,Nguyen-Thanh_2015a,Nguyen-Thanh_2017a}, however, based on the proposed view point, the implementation is quite different. In particular, when PDEs on manifolds from other application fields than shell mechanics are also of interest (e.g., when transport problems \cite{Demlow_2009a,Dziuk_1988a,Dziuk_2013a} or flow problems \cite{Fries_2018b,Jankuhn_2017a} on curved surfaces are considered), there is a unified and elegant way to handle this by computing surface gradients applied to finite element shape functions which simplifies the situation considerably. In that sense one may shift significant parts of the implementation needed for shells to the underlying finite element technology and recycle this in other situations where PDEs on surfaces are considered. \par

We summarize the advantages of the TDC-based formulation of Kirchhoff-Love shells: (1) The definition of the BVP does not need a parametrization of the surface (though it can also handle the classical situation where a parametrization is given), (2) the TDC-based formulation is also suitable for very recent finite element technologies such as CutFEM and TraceFEM (though the typical approach based on the Surface FEM or IGA is also possible and demonstrated herein), (3) the implementation is advantagous in finite element (FE) codes where other PDEs on manifolds are considered as well due to the split of FE technology and application. From a didactic point of view, it may also be advantageous that troubles with curvilinear coordinates (co- and contra-variance, Christoffel-symbols) are avoided in the TDC-based approach where surface operators and geometric quantities are expressed in tensor notation.
\revEnd

The outline of the paper is as follows: In \autoref{sec:pre}, important surface quantities are defined, and an introduction to the tangential differential calculus (TDC) is given. In \autoref{sec:shell}, the classical linear Kirchhoff-Love shell equations under static loading are recast in terms of the TDC. Stress resultants such as membrane forces, bending moments, transverse shear forces and corner forces are defined. In \autoref{sec:impl}, implementational aspects are considered. The element stiffness matrix and the resulting system of linear equations are shown. The implementation of boundary conditions based on Lagrange multipliers is outlined. Finally, in \autoref{sec:numres}, numerical results are presented. The first example is a flat shell embedded in $\mathbb{R}^3$, where an analytical solution is available. The second and third example are parts of popular benchmarks as proposed in \cite{Belytschko_1985a}. In the last example, a more general geometry without analytical solution or reference displacement is considered. The error is measured in the strong form of the equilibrium in order to verify the proposed approach and higher-order convergence rates are achieved.

%% file: Content/Preliminaries.tex
\section{Preliminaries}
\label{sec:pre}

\begin{figure}[h]
	\centering
	\subfloat[explicit]{\tikzfigforpaper{tikz/SurfExpl}{0.2}{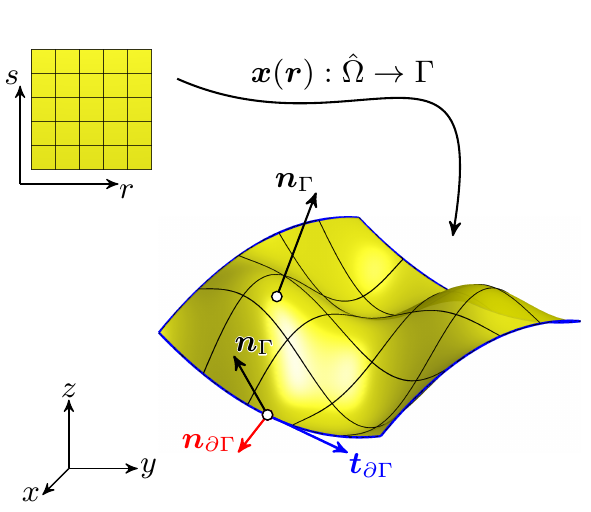} \label{fig:surfexpl}}
	\hfil
	\subfloat[implicit]{\tikzfigforpaper{tikz/SurfImpl}{0.2}{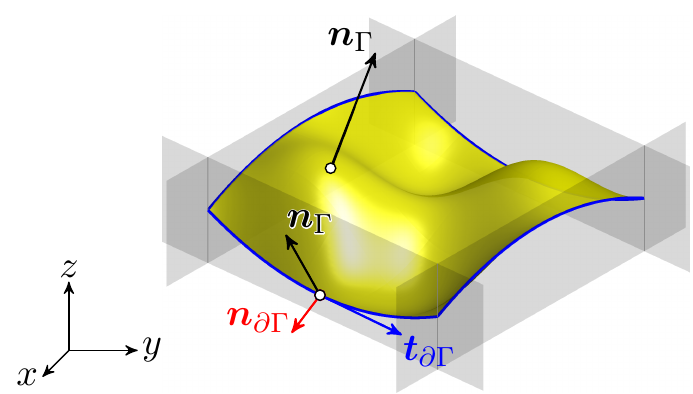}\label{fig:surfimpl}}
	\caption{Examples of bounded surfaces $\Gamma$ embedded in the physical space $\mathbb{R}^3$: (a) Explicitly defined surface with a map $\vek{x}(\vek{r})$, (b) Implicitly defined surface with a master level-function $\phi(\vek{x}) = 0$ (yellow) and slave level-set functions $\psi_i$ for the boundary definition (gray).}
	\label{fig:surf}
\end{figure}

Shells are geometrical objects, where one dimension is significantly smaller compared to the other two dimensions. In this case, the shell can be reduced to a surface $\Gamma$ embedded in the physical space $\mathbb{R}^3$. In particular, the surface is a manifold of codimension 1. Let the surface be possibly curved, sufficiently smooth, orientable, connected and bounded by $\p\Gamma$. There are two alternatives for defining the shell geometry. One is through a parametrization, i.e., a (bijective) mapping
\begin{align}
\vek{x}(\vek{r}) :\ &\hat{\Omega} \to \Gamma
\end{align} from the parameter space $\hat{\Omega} \subset \mathbb{R}^2$ to the real domain $\Gamma\subset \mathbb{R}^3$. The other approach is based on the level-set method. Then, a level-set function \revStart$\phi(\vek{x}) : \mathbb{R}^3 \to \mathbb{R}$ \revEnd with $\vek{x}\in\Omega\subset\mathbb{R}^3$ exists and the shell is implicitly given by 
\begin{align} \label{eq:implsurf}
\Gamma = \lbrace\vek{x}:\phi(\vek{x})=0 \quad \forall\  \vek{x}\in\Omega\rbrace\ .
\end{align}
Additional level-set functions may restrict the zero-isosurface to the desired, bounded shell as described in \cite{Fries_2017b}. In \autoref{fig:surfexpl} and (b) the two different approaches are schematically shown.
\revStart

The definition of the normal vector depends on whether the shell geometry is based on a parametrization or not. In the first case (cf.~\autoref{fig:surfexpl}), the shell geometry results from a map $\vek{x}(\vek{r})$. Then, the normal vector $\nG$ of the shell surface is determined by a cross-product of the columns of the Jacobi-matrix $\mat{J}(\vek{r})=\sfrac{\p\vek{x}}{\p\vek{r}}$. The resulting geometric quantities, surface operators, and models in this case are parametrization-based. \par
In the case where the shell geometry is implied by the zero-isosurface of a level-set function $\phi(\vek{x})$ (cf.~\autoref{fig:surfimpl}) and no parametrization is available, the normal vector may be determined by $\nG = \sfrac{\nabla\phi}{\Vert\nabla\phi\Vert}$. All resulting quantities including the BVP of the Kirchhoff-Love shell are parametrization-free in this case. Of course, when in the wake of discretizing the BVP, the Surface FEM is used for the approximation, then a surface mesh of the shell geometry is needed and the surface elements do imply a parametrization again. It was already mentioned above, that other numerical methods such as the TraceFEM and CutFEM do not rely on a surface mesh. In this case, the countinuous and discrete BVP for the shell are truly parametrization-free.\par

In addition to the normal vector on the surface, \revEnd along the boundary \revStart$\p\Gamma$ \revEnd there is an associated tangential vector $\vek{t}_{\p\Gamma} \in \mathbb{R}^3$ pointing in the direction of $\p\Gamma$ and a co-normal vector $\vek{n}_{\p\Gamma} = \vek{n}_\Gamma \times \vek{t}_{\p\Gamma} \in \mathbb{R}^3$ pointing \glqq outwards\grqq\ and being perpendicular to the boundary yet in the tangent plane of the surface $\Gamma$. For the proof of equivalence of both cases we refer to, e.g., \cite{Dziuk_2013a}.


\subsection{Tangential Differential Calculus}
\label{sec:tdc}

The TDC provides a framework to define differential operators avoiding the use of classical differential geometric methods based on local coordinate systems and Christoffel symbols. In the following, an overview of the operators and relations in the frame of the TDC are presented. For simplicity, we restrict ourselves to the case of surfaces embedded in the three dimensional space. However, the shown relations and definitions may be adopted to other situations accordingly (e.g.,~curved lines embedded in $2$D or $3$D). An introduction from a more mathematical point of view is given in \cite{Delfour_2011a,Gurtin_1975a,Jankuhn_2017a}.

\subsubsection*{Orthogonal projection operator $\mat{P}$}
\label{sec:tdcp}

The orthogonal projection operator or normal projector $\mat{P} \in \mathbb{R}^{3\times 3}$ is defined as
\begin{align}
\mat{P} = \mathbb{I} - \vek{n}_\Gamma \otimes \vek{n}_\Gamma\ .
\end{align}
The operator $\otimes$ is the dyadic product of two vectors. The normal projector $\mat{P}$ projects a vector $\vek{v}$ onto the tangent space $T_P\Gamma$ of the surface. Note that $\mat{P}$ is idempotent $(\mat{P} \cdot \mat{P} = \mat{P})$, symmetric $(\mat{P} = \mat{P}^\T)$ and obviously in the tangent space $T_P\Gamma$ of the surface, i.e.,~$\mat{P}\cdot\vek{n}_\Gamma = \nG^\T \cdot \mat{P} =~\vek{0})$.\par
The projection of a vector field $\vek{v} : \Gamma \to \mathbb{R}^3$ onto the tangent plane is defined by
\begin{align}
\vek{v}_t = \mat{P}\cdot\vek{v}\quad \in T_P\Gamma
\end{align}
where $\vek{v}_t$ is tangential, i.e.~$\vek{v}_t \cdot \nG = 0$. The double projection of a second-order tensor function $\ten{A}(\vek{x}) : \Gamma \to \mathbb{R}^{3\times3}$ leads to an in-plane tensor and is defined as
\begin{align}
\ten{A}_t = \mat{P} \cdot \ten{A} \cdot \mat{P} \quad \in T_P\Gamma\ ,
\end{align}
with the properties $\ten{A}_t = \mat{P} \cdot \ten{A}_t \cdot \mat{P}$ and $\ten{A}_t\cdot \nG = \nG^\T \cdot \ten{A}_t = \vek{0}$.

\subsubsection*{Tangential gradient of scalar functions}
The tangential gradient $\nabla_\Gamma$ of a scalar function $u : \Gamma \to \mathbb{R}$ on the manifold is defined as 
\begin{align}\label{eq:gradPu}
\nabla_\Gamma u(\vek{x}) = \mat{P}(\vek{x}) \cdot \nabla \tilde{u}(\vek{x})\ ,\ \nabla_\Gamma u(\vek{x}) \in \mathbb{R}^{3\times 1}\ ,\ \vek{x} \in \Gamma
\end{align}
where $\nabla$ is the standard gradient operator in the physical space and $\tilde{u}$ is a smooth extension of $u$ in a neighbourhood $\mathcal{U}$ of the manifold $\Gamma$. Alternatively, $\tilde{u}$ is given as a function in global coordinates $\tilde{u}(\vek{x}) : \mathbb{R}^3 \to \mathbb{R}$ and only evaluated at the manifold $\tilde{u}_{\rvert_\Gamma} = u$.\par
For parametrized surfaces defined by the map $\vek{x}(\vek{r})$, and a given scalar function $u(\vek{r}) : \hat{\Omega} \to \mathbb{R}$, the tangential gradient can be determined without explicitly computing an extension $\tilde{u}$ using
\begin{align}\label{eq:gradu}
\nabla_\Gamma u(\vek{x}(\vek{r})) = \mat{J}(\vek{r}) \cdot \mat{G}(\vek{r})^{-1} \cdot \gradR{u}(\vek{r})\ ,
\end{align}
with $\mat{J}(\vek{r}) = \sfrac{\p\vek{x}}{\p\vek{r}} \in \mathbb{R}^{3\times 2}$ being the Jacobi-matrix, $\mat{G} = \mat{J}^\T\cdot\mat{J}$ is the metric tensor or the first fundamental form and the operator $\gradR{}$ is the gradient with respect to the reference coordinates. The components of the tangential gradient are denoted by
\begin{align}
\nabla_\Gamma u = \begin{bmatrix}
\p^\Gamma_x u \\
\p^\Gamma_y u \\
\p^\Gamma_z u
\end{bmatrix}\ ,
\end{align}
representing first-order partial tangential derivatives. An important property of $\nabla_\Gamma u$ is that the tangential gradient of a scalar-valued function is in the tangent space of the surface $\nabla_\Gamma u \in T_P\Gamma$, i.e.,~$\nabla_\Gamma u \cdot \vek{n}_\Gamma = 0$. When using the Surface FEM to solve BVPs on surfaces, one may use \autoref{eq:gradu} to compute tangential gradients of the shape functions. \revStart If, on the other hand, TraceFEM or CutFEM is used, one may use \autoref{eq:gradPu}. \revEnd

\subsubsection*{Tangential gradient of vector-valued functions}

Consider a vector-valued function $\vek{v}(\vek{x}) : \Gamma \to \mathbb{R}^3$ and apply to each component of $\vek{v}$ the tangential gradient for scalars. This leads to the \emph{directional} gradient of $\vek{v}$ defined as 
\begin{align}\label{eq:gradvdir}
\gradGD \vek{v}(\vek{x}) = \gradGD \begin{bmatrix}
u(\vek{x}) \\
v(\vek{x}) \\
w(\vek{x})
\end{bmatrix} = \begin{bmatrix}
\p^\Gamma_x u & \p^\Gamma_y u & \p^\Gamma_z u \\
\p^\Gamma_x v & \p^\Gamma_y v & \p^\Gamma_z v \\
\p^\Gamma_x w & \p^\Gamma_y w & \p^\Gamma_z w
\end{bmatrix}\ .
\end{align}
Note that the directional gradient is \emph{not} in the tangent space of the surface, in general. A projection of the directional gradient to the tangent space leads to the \emph{covariant} gradient of $\vek{v}$ and is defined as 
\begin{align}
\gradGC \vek{v} = \mat{P} \cdot \gradGD \vek{v}\ ,
\end{align}
which is an in-plane tensor, i.e., $\gradGC\vek{v} \in T_P\Gamma$. The covariant gradient often appears in the modelling of physical phenomena on manifolds, i.e., in the governing equations. In contrast the directional gradient appears naturally in product rules or divergence theorems on manifolds.

 In the following, partial surface derivatives of scalar functions are denoted as $ \p^\Gamma_{x_i} u$ or $u_{,i}^\Gamma$ with $i = 1,\,2,\,3$. Partial surface derivatives of vector or tensor components are denoted as $v_{i,j}^{\t{dir}}$ for directional and $v_{i,j}^{\t{cov}}$ for covariant derivatives with $i,j = 1,\,2,\,3$.

\subsubsection*{Tangential gradient of tensor functions}
For a second-order tensor function $\mat{A}(\vek{x}) : \Gamma \to \mathbb{R}^{3\times 3}$, the partial directional gradient with respect to $x_i$ is defined as
\begin{align}
\nabla^{\text{dir}}_{\Gamma, i} \mat{A} = \dfrac{\p \mat{A}}{\p^\Gamma_{x_i}} = \begin{bmatrix}
\p^\Gamma_{x_i}A_{11} & \p^\Gamma_{x_i}A_{12} & \p^\Gamma_{x_i}A_{13} \\
\p^\Gamma_{x_i}A_{21} & \p^\Gamma_{x_i}A_{22} & \p^\Gamma_{x_i}A_{23} \\
\p^\Gamma_{x_i}A_{31} & \p^\Gamma_{x_i}A_{32} & \p^\Gamma_{x_i}A_{33}
\end{bmatrix} \quad \text{with: } i = 1,\,2,\,3\ .
\end{align}
The directional gradient of the tensor function is then defined as
\begin{align}
\gradGD \mat{A} = \left( \nabla^{\text{dir}}_{\Gamma, 1}\mat{A}\quad \nabla^{\text{dir}}_{\Gamma, 2}\mat{A}\quad \nabla^{\text{dir}}_{\Gamma, 3} \mat{A} \right)\ .
\end{align}

The covariant partial derivative is determined by projecting the partial directional derivative onto the tangent space
\begin{align}
\nabla^\text{cov}_{\Gamma, i} \mat{A} = \mat{P} \cdot \nabla^{\text{dir}}_{\Gamma, i} \mat{A} \cdot \mat{P}\ .
\end{align}

\subsubsection*{Second-order tangential derivatives}
Next, second-order derivatives of scalar functions are considered. The directional second order gradient of a scalar function $u$ is defined by
\begin{align}
\lbrace\mat{H}\mat{e}{^\t{dir}}\rbrace_{ij}(u(\vek{x})) = \p^{\Gamma,\,\t{dir}}_{x_j}\left(\p^\Gamma_{x_i} u(\vek{x})\right) = u_{,ji}^{\t{dir}}= \begin{bmatrix}
\p^\Gamma_{xx} u & \p^\Gamma_{yx} u & \p^\Gamma_{zx} u \\
\p^\Gamma_{xy} u & \p^\Gamma_{yy} u & \p^\Gamma_{zy} u \\
\p^\Gamma_{xz} u & \p^\Gamma_{yz} u &  \p^\Gamma_{zz} u
\end{bmatrix} = \gradGD\left(\gradG u(\vek{x})\right)
\end{align}
where $\mat{H}\mat{e}^{\t{dir}}$ is the tangential Hessian matrix which is not symmetric in the case of curved manifolds \cite{Delfour_2011a}, i.e., $u_{,ij}^{\t{dir}} \neq u_{,ji}^{\t{dir}}$. For the case of parametrized surfaces and a given scalar function in the reference space, the tangential Hessian-matrix can be determined by
\begin{align}
\begin{split}
\mat{H}\mat{e}^{\t{dir}}(u) &= \gradGD\left( \mat{Q} \cdot \gradR{u} \right) \\
&= \left[\mat{Q}_{,r} \cdot \gradR{u} \quad \mat{Q}_{,s} \cdot \gradR{u}\right] \cdot \mat{Q}^\T + \mat{Q} \cdot \gradR{\left(\gradR{u}\right)} \cdot \mat{Q}^\T
\end{split}
\end{align}
where, $\mat{Q} = \mat{J} \cdot \mat{G}^{-1}$, and $\mat{Q}_{,r_i}$ denotes the partial tangential derivative of $\mat{Q}$ with respect to $r_i$. The covariant counterpart is
\begin{align}
\mat{H}\mat{e}^\text{cov}(u) = \gradGC\left(\gradG u\right) = \mat{P} \cdot \gradGD \left(\gradG u\right) = \mat{P} \cdot \mat{H}\mat{e}^\text{dir}(u)\ .
\end{align}
In contrast to $\mat{H}\mat{e}^\text{dir}$, $\mat{H}\mat{e}^\text{cov}$ is symmetric and an in-plane tensor \cite{Walker_2015a}. In the special case of flat surfaces embedded in $\mathbb{R}^3$ the directional and covariant Hessian matrix are equal.

\subsubsection*{Tangential divergence operators}
The divergence operator of a vector-valued function $\vek{v}(\vek{x}) : \Gamma \to \mathbb{R}^3$ is given as
\begin{align}
\divG  \vek{v}(\vek{x}) = \text{tr}\left(\gradGD \vek{v}(\vek{x})\right) = \text{tr}\left(\gradGC \vek{v}(\vek{x})\right)\ ,
\end{align}
and the divergence of a matrix or tensor function $\mat{A}(\vek{x}) : \Gamma \to \mathbb{R}^{3\times 3}$, is
\begin{align}
\divG  \mat{A}(\vek{x}) = \begin{bmatrix}
\divG \left[ A_{11},\, A_{12},\, A_{13}\right] \\
\divG \left[ A_{21},\, A_{22},\, A_{23}\right] \\
\divG \left[ A_{31},\, A_{32},\, A_{33}\right]
\end{bmatrix}\ .
\end{align}
Note that $\divG\mat{A}$ is, in general, not a tangential vector. It would only be tangential if the surface is flat \emph{and} $\mat{A}$ is an in-plane tensor.

\subsubsection*{Weingarten map and curvature}
\begin{figure}[ht]
	\tikzfigforpaper{tikz/PrincipleCurvature}{0.15}{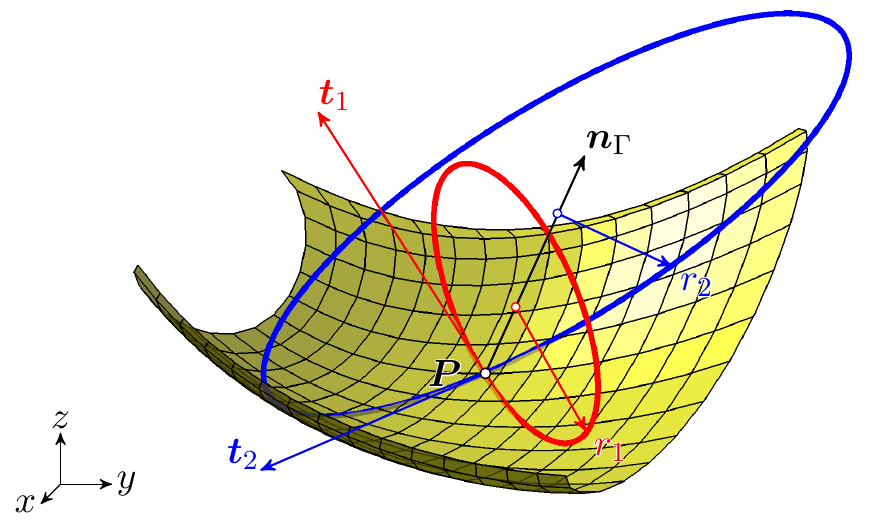}
	\caption{Osculating circles (blue, red) and eigenvectors $(\vek{t}_1,\,\vek{t}_2,\,\nG)$ of $\mat{H}$ at point $\vek{P}$ on a surface embedded in $\mathbb{R}^3$.}
	\label{fig:princurv}
\end{figure}

The Weingarten map as introduced in \cite{Jankuhn_2017a, Delfour_2011a} is defined as
\begin{align}
\mat{H} = \gradGD\vek{n}_\Gamma = \gradGC\vek{n}_\Gamma
\end{align}
and is related to the second fundamental form in differential geometry. The Weingarten map is a symmetric, in-plane tensor and its two non-zero eigenvalues are associated with the principal curvatures
\begin{align}
\kappa_{1,2} = -\,\t{eig}(\mat{H})\ . \label{eq:prcurv}
\end{align}
The minus in \autoref{eq:prcurv} is due to fact that the Weingarten map is defined with the \glqq outward\grqq\ unit normal vector instead of the \glqq inward\grqq\ unit normal vector, which leads to positive curvatures of a sphere. The third eigenvalue is zero, because $\mat{H}$ is an in-plane tensor. The corresponding eigenvectors $\vek{t}_1,\,\vek{t}_2$ and $\nG$ are perpendicular as $\mat{H}$ is symmetric. In \autoref{fig:princurv}, the osculating circles with the radii $r_i = \sfrac{1}{\kappa_i}$ and the eigenvectors at a point $\vek{P}$ are shown.

The Gau\ss\ curvature is defined as the product of the principal curvatures $K = \prod_{i=1}^{2}\kappa_i$ and the mean curvature is introduced as $\varkappa = \kappa_1 + \kappa_2 = \t{tr}(\mat{H})$.

\subsubsection*{Divergence theorems in terms of tangential operators}

The divergence theorem or Green's formula for a scalar function $f \in C^1(\Gamma)$ and a vector valued function $\vek{v} \in C^1(\Gamma)^3$ are defined as in \cite{Delfour_1996a, Delfour_2011a} 
\begin{align}\label{eq:greenvec}
\begin{split}
\int_\Gamma f \cdot \divG  \vek{v}\ \d \Gamma = -&\int_\Gamma \nabla_\Gamma f \cdot \vek{v}\ \d \Gamma + \int_\Gamma \varkappa f \left(\vek{v} \cdot \vek{n}_\Gamma\right)\ \d \Gamma + \\
&\int_{\p\Gamma} f \vek{v} \cdot \vek{n}_{\p\Gamma}\ \d\p\Gamma\ .
\end{split}
\end{align}
The term with the mean curvature $\varkappa$ is vanishing if the vector $\vek{v}$ is tangential, then $\vek{v}\cdot \vek{n}_\Gamma=0$. In extension to \autoref{eq:greenvec}, Green's formula for second order tensor functions $\mat{A} \in C^1(\Gamma)^{3\times 3}$, is
\begin{align}
\begin{split}\label{eq:greenten}
\int_\Gamma \vek{v} \cdot \divG  \mat{A}\ \d \Gamma = -&\int_\Gamma \gradGD \vek{v} : \mat{A}\ \d \Gamma + \int_\Gamma \varkappa\, \vek{v} \cdot \left(\mat{A} \cdot \vek{n}_\Gamma\right)\ \d \Gamma + \\ &\int_{\p\Gamma} \vek{v} \cdot \left(\mat{A} \cdot \vek{n}_{\p\Gamma}\right)\ \d \p\Gamma
\end{split}
\end{align}
where $\gradGD \vek{v} : \mat{A} = \text{tr}(\gradGD \vek{v} \cdot \mat{A}^\T)$. In the case of in-plane tensors, e.g.,~$\mat{A}_t = \mat{P}\cdot\mat{A}_t\cdot\mat{P}$, the term with the mean curvature $\varkappa$ vanishes due to $\mat{A}_t\cdot\vek{n}_\Gamma = \vek{0}$ and we also have $\gradGD \vek{v} : \mat{A}_t = \gradGC \vek{v} : \mat{A}_t$.

%% file: Content/Methodology.tex
\section{The shell equations}
\label{sec:shell}

In this section, we derive the linear Kirchhoff-Love shell theory in the frame of tangential operators based on a global Cartesian coordinate system. We restrict ourselves to infinitesimal deformations, which means that the reference and spatial configuration are indistinguishable. Furthermore, a linear elastic material governed by Hooke's law is assumed. As usual in the Kirchhoff-Love shell theory, the transverse shear strains and the change of curvature in the material law are neglected, which restricts the model to thin shells $(t\kappa_{\max} \ll 1)$.\par

With these assumptions, an analytical pre-integration with respect to the thickness leads to stress resultants such as normal forces and bending moments. The equilibrium in strong form is then expressed in terms of the stress resultants. Finally, the transverse shear forces may be identified via equilibrium considerations.\par

\subsection{Kinematics}
\label{sec:klshellkin}
\begin{figure}[h]\centering
	\tikzfigforpaper{tikz/DisplacementVector}{0.2}{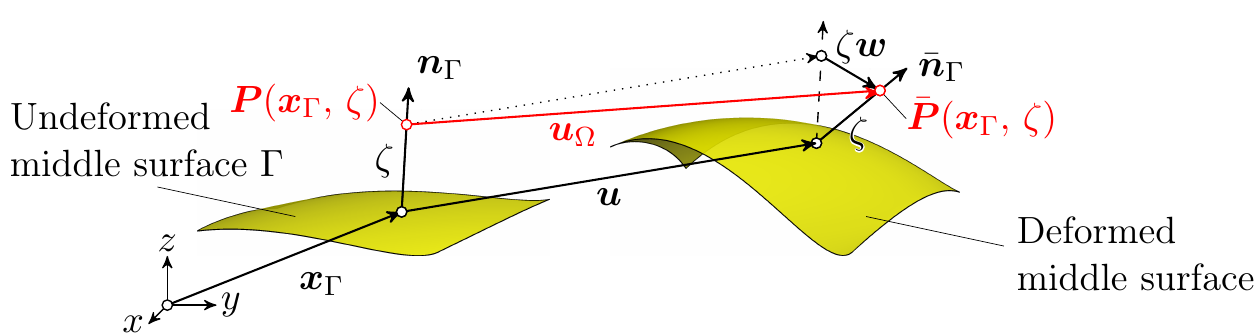}
	\caption{Displacements $\vek{u}_\Omega,\,\vek{u}$ and $\vek{w}$ of the shell.}
	\label{fig:dispu}
\end{figure}

The middle surface $\Gamma$ of the shell is a sufficiently smooth manifold embedded in the physical space $\mathbb{R}^3$. A point on the middle surface is denoted as $\vek{x}_\Gamma \in \Gamma \subset \mathbb{R}^3$ and may be obtained explicitly or implicitly, see \autoref{sec:pre}. With the unit-normal vector $\nG$ a point in the domain of the shell $\Omega$ of thickness $t$ is defined by
\begin{align}
\vek{x} = \vek{x}_\Gamma + \zeta\nG
\end{align}
with $\zeta$ being the thickness parameter and $\vert\zeta\vert \le \sfrac{t}{2}$. Alternatively, if the middle surface is defined implicitly with a signed distance function $\phi(\vek{x})$ the domain of the shell $\Omega$ is defined by 
\begin{align}
\Omega = \left\lbrace \vek{x} \in \mathbb{R}^3 : \vert\phi(\vek{x})\vert \le \frac{t}{2} \right\rbrace\ .
\end{align}
In this case the middle surface $\Gamma$ is the zero-isosurface of $\phi(\vek{x})$, see \autoref{eq:implsurf}. The displacement field $\vek{u}_\Omega$ of a point $\vek{P}(\vek{x}_\Gamma,\,\zeta)$ in the shell continuum $\Omega$ takes the form
\begin{align}
\vek{u}_\Omega(\vek{x}_\Gamma,\,\zeta) = \vek{u}(\vek{x}_\Gamma) + \zeta\vek{w}(\vek{x}_\Gamma)
\end{align}
with $\vek{u}(\vek{x}_\Gamma) = [u,\,v,\,w]^\T$ being the displacement field of the middle surface and $\vek{w}(\vek{x}_\Gamma)$ being the difference vector, as illustrated in \autoref{fig:dispu}.

Without transverse shear strains, the difference vector $\vek{w}$ expressed in terms of TDC is defined as in \cite{Delfour_1996a}
\begin{align} \label{eq:diffvec}
\vek{w}(\vek{x}_\Gamma) &= - \left[\gradGD{\vek{u}} + (\gradGD{\vek{u}})^\T\right]\cdot \nG = \mat{H}\cdot\vek{u} - \gradG{(\vek{u}\cdot\nG)}\ .
\end{align}
As readily seen in the equation above, the difference vector $\vek{w}$ is tangential. Alternatively, the difference vector $\vek{w}$ may also be re-written in terms of partial tangential derivatives of $\vek{u}$ and the normal vector $\nG$ 
\begin{align}
\vek{w}(\vek{x}_\Gamma) &= \mat{H}\cdot\vek{u} - \gradG{(\vek{u}\cdot\nG)} = -\begin{bmatrix}
\vek{u}_{,x}^\t{dir} \cdot \nG \\[.25cm]
\vek{u}_{,y}^\t{dir} \cdot \nG \\[.25cm]
\vek{u}_{,z}^\t{dir} \cdot \nG
\end{bmatrix}\ .
\end{align}
Consequently, the displacement field of the shell continuum is only a function of the middle surface displacement $\vek{u}$, the unit normal vector $\nG$ and the thickness parameter $\zeta$.\par

The linearised, in-plane strain tensor $\ten{\varepsilon}_\Gamma$ is defined by the symmetric part of the directional gradient of the displacement field $\vek{u}_\Omega$, projected with $\mat{P}$ \cite{Hansbo_2014a}
\begin{align}
\begin{split}
\ten{\varepsilon}_\Gamma (\vek{x}_\Gamma,\,\zeta) &= \mat{P}\cdot\dfrac{1}{2}\left[\gradGD{\vek{u}_\Omega} + (\gradGD{\vek{u}_\Omega})^\T\right]\cdot \mat{P} = \mat{P}\cdot \ten{\varepsilon}_\Gamma^\t{dir}\cdot \mat{P}\label{eq:epsco}\\
 &= \dfrac{1}{2}\left[\gradGC{\vek{u}_\Omega} + (\gradGC{\vek{u}_\Omega})^\T\right]\ .
\end{split}
\end{align}
Finally, the whole strain tensor may be split into a membrane and bending part, as usual in the classical theory
\begin{align}
\ten{\varepsilon}_\Gamma &=  \ten{\varepsilon}_{\Gamma,\t{M}}(\vek{u}) + \zeta\ten{\varepsilon}_{\Gamma,\t{B}}(\vek{w})\ , \label{eq:eps}
\intertext{with}
\ten{\varepsilon}_{\Gamma,\t{M}}   &= \dfrac{1}{2}(\gradGC{\vek{u}} + (\gradGC{\vek{u}})^\T)\ , \nonumber\\
\ten{\varepsilon}_{\Gamma,\t{B}}  &= -\begin{bmatrix}
\vek{u}_{,xx}^{\t{cov}} \cdot \nG & \vek{u}_{,yx}^{\t{cov}} \cdot \nG & \vek{u}_{,zx}^{\t{cov}} \cdot \nG \\
 & \vek{u}_{,yy}^{\t{cov}} \cdot \nG & \vek{u}_{,zy}^{\t{cov}} \cdot \nG \\
 \t{sym} & & \vek{u}_{,zz}^{\t{cov}} \cdot \nG
\end{bmatrix}\ . \nonumber
\end{align}
Note that in the linearised bending strain tensor $\ten{\varepsilon}_{\Gamma,\t{B}}$, the term $(\gradGD{\vek{u}})^\T \cdot \mat{H}$ is neglected as in classical theory \cite[Remark 2.2]{Simo_1989b} or \cite{Wempner_2002a}. The resulting membrane and bending strain in \autoref{eq:eps} are equivalent compared to the classical theory, e.g., \cite{Basar_1985a}. In the case of flat shell structures as considered in \cite{Hansbo_2017a} the membrane strain is only a function of the tangential displacement $\vek{u}_t = \mat{P}\cdot\vek{u}$ and the bending strain only depends on the normal displacement $u_n = \vek{u}\cdot\nG$, which simplifies the whole kinematic significantly. Moreover, the normal vector $\nG$ is then constant and the difference vector simplifies to $\vek{w}(\vek{x}_\Gamma) = -\gradG{u_n}$.

\subsection{Constitutive Equation}
\label{sec:klshellconst}

As already mentioned above, the shell is assumed to be linear elastic and, as usual for thin structures, plane stress is presumed. The in-plane stress tensor $\ten{\sigma}_\Gamma$ is defined as
\begin{align}
\ten{\sigma}_{\Gamma}(\vek{x}_\Gamma,\,\zeta) &= \mat{P}\cdot\left[2\mu\ten{\varepsilon}_\Gamma + \lambda\t{tr}(\ten{\varepsilon}_\Gamma)\mathbb{I}\right] \cdot \mat{P} \\
&=\mat{P}\cdot\left[2\mu\ten{\varepsilon}_\Gamma^\t{dir} + \lambda\t{tr}(\ten{\varepsilon}^\t{dir}_\Gamma)\mathbb{I}\right] \cdot \mat{P} \label{eq:stressdir}
\end{align}
where $\mu = \frac{E}{2(1+\nu)}$ and $\lambda = \frac{E\nu}{(1-\nu^2)}$ are the Lam\'e constants and $\ten{\varepsilon}_\Gamma^\t{dir}$ is the directional strain tensor from \autoref{eq:epsco}. With this identity the in-plane stress tensor can be computed only with the directional strain tensor
\begin{align*}
\ten{\varepsilon}_\Gamma^\t{dir} &=  \ten{\varepsilon}_{\Gamma,\t{M}}^\t{dir}(\vek{u}) + \zeta\ten{\varepsilon}_{\Gamma,\t{B}}^\t{dir}(\vek{w})\ , \label{eq:epsdir}
\intertext{with}
\ten{\varepsilon}_{\Gamma,\t{M}}^\t{dir}   &= \dfrac{1}{2}(\gradGD{\vek{u}} + (\gradGD{\vek{u}})^\T)\ , \nonumber\\
\ten{\varepsilon}_{\Gamma,\t{B}}^\t{dir}  &= -\begin{bmatrix}
\vek{u}_{,xx}^{\t{dir}} \cdot \nG & \frac{1}{2}(\vek{u}_{,yx}^{\t{dir}} + \vek{u}_{,xy}^{\t{dir}}) \cdot \nG & \frac{1}{2}(\vek{u}_{,zx}^{\t{dir}} + \vek{u}_{,xz}^{\t{dir}}) \cdot \nG \\
& \vek{u}_{,yy}^{\t{dir}} \cdot \nG & \frac{1}{2}(\vek{u}_{,zy}^{\t{dir}} + \vek{u}_{,yz}^{\t{dir}}) \cdot \nG \\
\t{sym} & & \vek{u}_{,zz}^{\t{dir}} \cdot \nG
\end{bmatrix}, \nonumber
\end{align*}

which is from an implementational point of view an advantage, because covariant derivatives are not needed explicitly. In comparison to the classical theory, the in-plane stress tensor expressed in terms of TDC does not require the computation of the metric coefficients in the material law. Therefore, the resulting stress tensor does not hinge on a parametrization of the middle surface and shell analysis on implicitly defined surfaces is enabled.

\subsubsection{Stress resultants}
\label{sec:stressres}
The stress tensor is only a function of the middle surface displacement vector $\vek{u}$, the difference vector $\vek{w}(\vek{u})$ and the thickness parameter $\zeta$. This enables an analytical pre-integration with respect to the thickness and stress resultants can be identified. The following quantities are equivalent to the stress resultants in the classical theory \cite{Simo_1989b, Basar_1985a}, but they are expressed in terms of the TDC using a global Cartesian coordinate system. \par

The moment tensor $\mat{m}_\Gamma$ is defined as
\begin{align}
\mat{m}_\Gamma &= \int_{-\sfrac{t}{2}}^{\sfrac{t}{2}} \zeta \ten{\sigma}_\Gamma(\vek{u},\,\zeta)\ \d\zeta = \dfrac{t^3}{12} \ten{\sigma}_\Gamma(\ten{\varepsilon}_{\Gamma,\t{B}}) = \mat{P}\cdot {\mat{m}}^\t{dir}_\Gamma \cdot\mat{P}\ ,
\intertext{with}
{\mat{m}}^\t{dir}_\Gamma &= - D_{\t{B}}\begin{bmatrix}
(\vek{u}^\t{dir}_{,xx} + \nu\vek{u}^\t{dir}_{,yy}+\nu\vek{u}^\t{dir}_{,zz})\cdot\nG& \frac{1-\nu}{2}(\vek{u}^\t{dir}_{,yx}+\vek{u}^\t{dir}_{,xy})\cdot\nG & \frac{1-\nu}{2}(\vek{u}^\t{dir}_{,zx}+\vek{u}^\t{dir}_{,xz})\cdot\nG \\[.25cm]
& (\vek{u}^\t{dir}_{,yy} + \nu\vek{u}^\t{dir}_{,xx}+\nu\vek{u}^\t{dir}_{,zz})\cdot\nG & \frac{1-\nu}{2}(\vek{u}^\t{dir}_{,zy}+\vek{u}^\t{dir}_{,yz})\cdot\nG \\[.25cm]
\t{sym.}&  &(\vek{u}^\t{dir}_{,zz} + \nu\vek{u}^\t{dir}_{,xx}+\nu\vek{u}^\t{dir}_{,yy})\cdot\nG\nonumber
\end{bmatrix}
\end{align}
where $D_{\t{B}} = \frac{Et^3}{12(1-\nu^2)}$ is the flexural rigidity of the shell. The moment tensor $\mat{m}_\Gamma$ is symmetric and an in-plane tensor. Therefore, one of the three eigenvalues is zero and the two non-zero eigenvalues of $\mat{m}_\Gamma$ are the principal bending moments $m_{1}$ and $m_{2}$.  The principal moments are in agreement with the eigenvalues of the moment tensor in the classical setting, see \cite{Basar_1985a}. For the effective normal force tensor $\tilde{\mat{n}}_\Gamma$ we have
\begin{align}
\tilde{\mat{n}}_\Gamma &= \int_{-\sfrac{t}{2}}^{\sfrac{t}{2}} \ten{\sigma}_\Gamma(\vek{u},\,\zeta)\ \d\zeta = t \ten{\sigma}_\Gamma(\ten{\varepsilon}_{\Gamma,\t{M}}) = \mat{P}\cdot {\mat{n}}^\t{dir}_\Gamma \cdot\mat{P}\ ,
\intertext{with}
{\mat{n}}^\t{dir}_\Gamma &= \dfrac{Et}{1-\nu^2}\begin{bmatrix}
u_{,x}^\t{dir} + \nu(v_{,y}^\t{dir}+w_{,z}^\t{dir}) & \frac{1-\nu}{2}(u_{,y}^\t{dir}+v_{,x}^\t{dir}) & \frac{1-\nu}{2}(u_{,z}^\t{dir}+w_{,x}^\t{dir}) \\[.25cm]
& v_{,y}^\t{dir} + \nu(u_{,x}^\t{dir}+w_{,z}^\t{dir}) & \frac{1-\nu}{2}(v_{,z}^\t{dir}+w_{,y}^\t{dir}) \\[.25cm]
\t{sym}&  & w_{,z}^\t{dir} + \nu(u_{,x}^\t{dir}+v_{,y}^\t{dir})\nonumber
\end{bmatrix}\ .
\end{align}
Similar to the moment tensor, the two non-zero eigenvalues of $\tilde{\mat{n}}_\Gamma$ are in agreement with the effective normal force tensor expressed in local coordinates. Note that for curved shells this tensor is \emph{not} the physical normal force tensor. This tensor only appears in the variational formulation, see \autoref{sec:impl}. The physical normal force tensor $\mat{n}^{\t{real}}_\Gamma$ is defined by 
\begin{align} \label{eq:nreal}
\mat{n}^{\t{real}}_\Gamma = \tilde{\mat{n}}_\Gamma + \mat{H}\cdot\mat{m}_\Gamma
\end{align}
and is, in general, not symmetric and also has one zero eigenvalue. The occurrence of the zero eigenvalues in $\mat{m}_\Gamma$, $\tilde{\mat{n}}_\Gamma$ and $\mat{n}^{\t{real}}_\Gamma$ is due to fact that these tensors are in-plane tensors, i.e.~$\mat{m}_\Gamma \cdot \nG = \nG^\T \cdot \mat{m}_\Gamma = \vek{0}$\ . The normal vector $\nG$ is the corresponding eigenvector to the zero eigenvalue and the other two eigenvectors are tangential.

\subsection{Equilibrium}
\label{sec:klshelleq}

Based on the stress resultants from above, one obtains the equilibrium for a curved shell in strong form as
\begin{align}	\label{eq:sf}
\divG{\tilde{\mat{n}}_\Gamma} + \nG\divG{(\mat{P}\cdot\divG{\mat{m}_\Gamma})} + 2\mat{H}\cdot\divG{\mat{m}_\Gamma} + [\p x_i^\Gamma\mat{H}]_{jk}[\mat{m}_\Gamma]_{ki} = -\vek{f}\ ,
\end{align}
with $\vek{f}$ being the load vector per area on the middle surface $\Gamma$. A summation over the indices $i,\,k = 1,\,2,\,3$ has to be performed. The obtained equilibrium does not rely on a parametrization of the middle surface but is, otherwise, equivalent to the equilibrium in local coordinates \cite{Basar_1985a, Wempner_2002a}. From this point of view, the reformulation of the linear Kirchhoff-Love shell equations in terms of the TDC may be seen as a generalization, because the requirement of a parametrized middle surface is circumvented. With boundary conditions, as shown in detail in \autoref{sec:klshellbc}, the complete fourth-order boundary value problem (BVP) is defined.

\par Based on the equilibrium in \autoref{eq:sf}, the transverse shear force vector $\vek{q}$ is defined as
\begin{align}
\vek{q} = \mat{P}\cdot\divG{\mat{m}_\Gamma}\ .
\end{align}
Note that in the special case of \emph{flat} Kirchhoff-Love structures embedded in $\mathbb{R}^3$ the divergence of an in-plane tensor is a tangential vector, as already mentioned in \autoref{sec:tdc}. Therefore, the definition of the transverse shear force vector in \cite{Hansbo_2017a} is in agreement with the obtained transverse shear force vector herein.

\subsubsection{Equilibrium in weak form}
\label{sec:klshellwf}

The equilibrium in strong form is converted to a weak form by multiplying \autoref{eq:sf} with a suitable test function $\vek{v}$ and integrating over the domain, leading to
\begin{align}
\begin{split}
-\int_{\Gamma} \vek{v}\cdot \lbrace\divG{\tilde{\mat{n}}_\Gamma} + \nG\divG{(\mat{P}\cdot\divG{\mat{m}_\Gamma})} + 2\mat{H}\cdot\divG{\mat{m}_\Gamma} + &[\p x_i^\Gamma\mat{H}]_{jk}[\mat{m}_\Gamma]_{ki}\rbrace\ \d\Gamma =\\
& =\int_{\Gamma} \vek{v}\cdot\vek{f}\ \d\Gamma\ .
\end{split}
\end{align}
With Green's formula from \autoref{sec:tdcp}, we introduce the continuous weak form of the equilibrium:\par
Find $\vek{u} \in \mathcal{V}:\Gamma \to \mathbb{R}^3$ such that
\begin{align}
a(\vek{u},\,\vek{v}) &= \langle \vek{F},\,\vek{v}\rangle \quad \forall\ \vek{v} \in \mathcal{V}_0\ , \label{eq:wf}
\intertext{with}
a(\vek{u},\,\vek{v}) &= \int_{\Gamma} \gradGD{\vek{v}}:\tilde{\mat{n}}_{\Gamma} -
\ten{\varepsilon}_{\Gamma,\t{B}}^\t{dir}(\vek{v}):\mat{m}_{\Gamma}\ \d \Gamma \nonumber ,\\
\begin{split}
\langle \vek{F},\,\vek{v}\rangle &= \int_{\Gamma} \vek{f} \cdot \vek{v}\ \d \Gamma - \int_{\p\Gamma_\t{N}}
\gradGD{(\vek{v}\cdot\nG)} \cdot (\mat{m}_\Gamma \cdot \nCo) - 2(\mat{H\cdot\vek{v}}) \cdot (\mat{m}_\Gamma \cdot \nCo) -
 \\
\phantom{\langle \vek{F},\,\vek{v}\rangle} & \hspace{3.8cm} \vek{v}\cdot(\tilde{\mat{n}}_\Gamma\cdot \nCo) - (\vek{v}\cdot\nG)\left(\mat{P}\cdot\divG{\mat{m}_\Gamma}\cdot\nCo\right)  \ \d s\ . \nonumber\end{split}
\end{align}
The corresponding function spaces are
\begin{align}
\mathcal{V} &= \lbrace \vek{u} : \Gamma \to \mathbb{R}^3\ \vert\ \vek{u} \in \mathcal{H}^1(\Gamma)^3 : \vek{u}_{,ji} \cdot \nG \in \mathcal{L}^2(\Gamma)^{3} \rbrace \label{eq:funspaceV}\\
\mathcal{V}_0 &= \lbrace \vek{v} \in \mathcal{V}(\Gamma) : \vek{v}{\rvert_{\p\Gamma_\t{D}}} = \vek{0} \rbrace\ 
\end{align}
where $\p\Gamma_\t{D}$ is the Dirichlet boundary and $\p\Gamma_\t{N}$ is the Neumann boundary. The advantage of this procedure is that the boundary terms naturally occur and directly allow to consider for mechanically meaningful boundary conditions. 

\subsubsection{Boundary conditions}
\label{sec:klshellbc}

As well known in the classical Kirchhoff-Love shell theory, special attention needs to be paid to the boundary conditions. In the following, the boundary terms of the weak form in \autoref{eq:wf} are rearranged in order to derive the effective boundary forces.\par

\begin{figure}[h]\centering
	\subfloat[rotations]{\tikzfigforpaper{tikz/BCRotVec}{0.2}{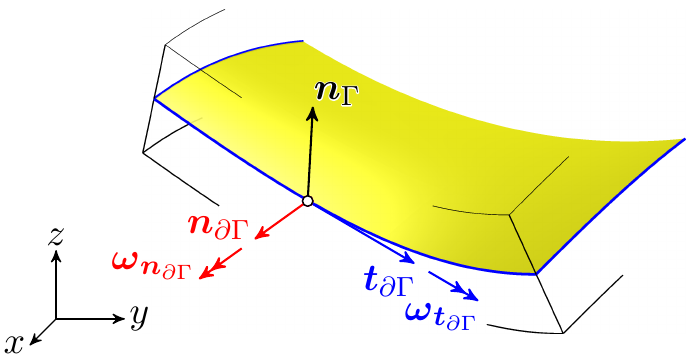}\label{fig:bcrot}}
	\hfil
	\subfloat[forces and moments]{\tikzfigforpaper{tikz/BCNM}{0.2}{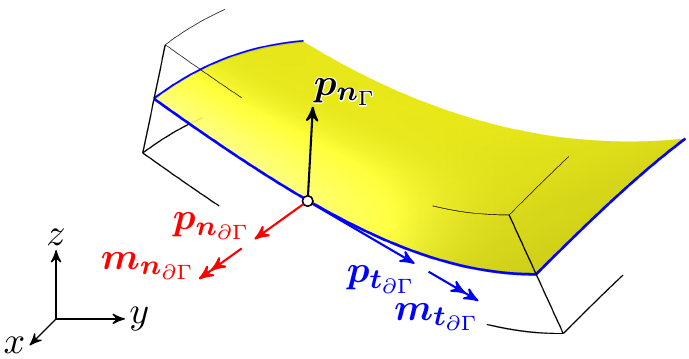}\label{fig:bcnm}}	
	\caption{Decomposition of the difference vector $\vek{w}$, in-plane normal forces $\mat{n}_\Gamma^{\t{real}}\cdot\nCo$ and bending moments $\mat{m}_\Gamma\cdot\nCo$  along the boundary $\p\Gamma$ in terms of $\vek{t}_{\p\Gamma}$ and $\nCo$: (a) Rotations at the boundary, (b) normal force tensor and bending moments at the boundary.}	
	\label{fig:bcdecomp}
\end{figure}

Using Eqs.~\eqref{eq:nreal} and \eqref{eq:diffvec}, we have
\begin{align}
-&\int_{\p\Gamma_\t{N}}
\gradGD{(\vek{v}\cdot\nG)} \cdot (\mat{m}_\Gamma \cdot \nCo) - 2(\mat{H\cdot\vek{v}}) \cdot (\mat{m}_\Gamma \cdot \nCo) - \vek{v}\cdot(\tilde{\mat{n}}_\Gamma\cdot \nCo) \nonumber
\\
&\phantom{\int_{\p\Gamma_\t{N}}\ }  - (\vek{v}\cdot\nG)\left(\mat{P}\cdot\divG{\mat{m}_\Gamma}\cdot\nCo\right)  \ \d s\ = \nonumber\\[.25cm] 
&\int_{\p\Gamma_\t{N}} \vek{v}\cdot(\mat{n}_\Gamma^{\t{real}}\cdot\nCo) + \vek{w}(\vek{v})\cdot(\mat{m}_\Gamma\cdot\nCo) + (\vek{v}\cdot\nG)\cdot\left(\mat{P}\cdot\divG{\mat{m}_\Gamma}\cdot\nCo\right)  \ \d s\ . \label{eq:bctermslong}
\end{align}
As already mentioned above, the difference vector $\vek{w}$ is a tangential vector. Consequently, the difference vector at the boundary may be expressed in terms of the tangential vectors $\vek{t}_{\p\Gamma}$ and $\nCo$
\begin{align}
\vek{w}(\vek{v}) &= \underset{ \mbox{$ \omega_{\vek{t}_{\p\Gamma}}$} }{\underbrace{\left[\mat{H}\cdot\vek{v} - \gradG{(\vek{v}\cdot\nG)}\right]\cdot\nCo}}\, \vek{t}_{\p\Gamma} + \underset{ \mbox{$ \omega_{\nCo}$} }{\underbrace{\left[\mat{H}\cdot\vek{v} - \gradG{(\vek{v}\cdot\nG)}\right]\cdot\vek{t}_{\p\Gamma}}}\, \nCo
\end{align}
where $\vek{\omega}_{\vek{t}_{\p\Gamma}} = \omega_{\vek{t}_{\p\Gamma}}\,\vek{t}_{\p\Gamma}$ may be interpreted as rotation along the boundary and $\vek{\omega}_{\nCo} = \omega_{\nCo}\, \nCo$ is the rotation in co-normal direction, when the test function $\vek{v}$ is interpreted as a displacement, see \autoref{fig:bcrot}. Analogously to the difference vector, the expressions $\mat{n}_\Gamma^{\t{real}}\cdot\nCo$ and $\mat{m}_\Gamma\cdot\nCo$ in \autoref{eq:bctermslong} are decomposed in a similar manner
\begin{align}
\mat{n}_\Gamma^{\t{real}}\cdot\nCo &= \underset{\mbox{$ p_{\vek{t}_{\p\Gamma}}$} }{\underbrace{(\mat{n}_\Gamma^{\t{real}}\cdot\nCo)\cdot\vek{t}_{\p\Gamma}}}\,\vek{t}_{\p\Gamma} + \underset{\mbox{$ p_{\nCo}$} }{\underbrace{(\mat{n}_\Gamma^{\t{real}}\cdot\nCo)\cdot\nCo}}\,\nCo\\[.25cm]
\mat{m}_\Gamma\cdot\nCo &= \underset{\mbox{$ m_{\vek{t}_{\p\Gamma}}$} }{\underbrace{(\mat{m}_\Gamma\cdot\nCo)\cdot\nCo}}\,\vek{t}_{\p\Gamma} + \underset{\mbox{$ m_{\nCo}$} }{\underbrace{(\mat{m}_\Gamma\cdot\nCo)\cdot\vek{t}_{\p\Gamma}}}\,\nCo\ .
\end{align}
Next, the term $p_{\nG} = \mat{P}\cdot\divG{\mat{m}_\Gamma}\cdot\nCo$ represents the resultant force in normal direction. In \autoref{fig:bcnm} the forces and bending moments along a curved boundary are illustrated.

Inserting these expressions in \autoref{eq:bctermslong}, the integral along the Neumann boundary simplifies to
\begin{align}
\int_{\p\Gamma_\t{N}} \vek{v}\cdot \left( {p}_{\tB}\tB + {p}_{\nCo}\nCo + {p}_{\nG}\nG\right) + \omega_{\tB}m_{\tB} + \omega_{\nCo}m_{\nCo}\ \d s\ .
\end{align}

As discussed in detail, e.g., in \cite{Basar_1985a}, the rotation in co-normal direction $\omega_{\nCo}$ is already prescribed with $\vek{v}\rvert_{\p\Gamma}$. Therefore, the term $\omega_{\nCo}m_{\nCo}$ is expanded and with integration by parts we obtain
\begin{align}
\begin{split}
\int_{\p\Gamma_\t{N}}\omega_{\nCo}m_{\nCo}\ \d s &= \int_{\p\Gamma_\t{N}} -\gradG{(\vek{v}\cdot\nG)}\cdot\tB\, m_{\nCo} + \mat{H}\cdot\vek{v}\cdot\tB\, m_{\nCo}\  \d s \\
&= \int_{\p\Gamma_\t{N}}  (\vek{v}\cdot\nG)\cdot\left(\gradG{m_{\nCo}}\cdot\tB\right) + \mat{H}\cdot\vek{v}\cdot\tB\, m_{\nCo}\ \d s \\
&\phantom{=\ } - (\vek{v}\cdot\nG)m_{\nCo} \big|_{+C}^{-C}\ 
\end{split}
\end{align}
where ${+C}$ and $-C$ are points close at a corner $C$. The new boundary term are the Kirchhoff forces or corner forces. Note that if the boundary of the shell is smooth, the corner forces vanish. Finally, the integral over the Neumann boundary \revStart in \autoref{eq:wf} \revEnd is expressed in terms of the well-known effective boundary forces and the bending moment along the boundary
\begin{align} \label{eq:bctermshort}
&\int_{\p\Gamma_\t{N}} \vek{v} \cdot \left( \tilde{p}_{\tB}\tB + \tilde{p}_{\nCo}\nCo + \tilde{p}_{\nG}\nG\right) + \omega_{\tB}m_{\tB}\ \d s - (\vek{v}\cdot\nG)m_{\nCo} \big|_{+C}^{-C} \\
\intertext{with:}
\tilde{p}_{\tB} &=  {p}_{\tB} + \left(\mat{H}\cdot\tB\right)\cdot\tB\,m_{\nCo} \\
\tilde{p}_{\nCo} &=  {p}_{\nCo} + \left(\mat{H}\cdot\tB\right)\cdot\nCo\,m_{\nCo} \\
\tilde{p}_{\nG} &=  {p}_{\nG} + \gradG{m_{\nCo}}\cdot\tB\ .
\end{align}
The obtained effective boundary forces and moments are in agreement with the given quantities in local coordinates \cite{Basar_1985a, Wempner_2002a}. The prescribeable boundary conditions are the conjugated displacements and rotations to the effective forces and moments at the boundary
\begin{align*}
&\tilde{p}_{\tB} = {p}_{\tB} + \left(\mat{H}\cdot\tB\right)\cdot\tB\,m_{\nCo} && \Longleftrightarrow &&\vek{u}\cdot\tB = u_{\tB}\ , \\
&\tilde{p}_{\nCo}= {p}_{\nCo} + \left(\mat{H}\cdot\tB\right)\cdot\nCo\,m_{\nCo}&& \Longleftrightarrow&&\vek{u}\cdot\nCo = u_{\nCo}\ , \\
&\tilde{p}_{\nG} =  {p}_{\nG} + \gradG{m_{\nCo}}\cdot\tB &&\Longleftrightarrow &&\vek{u}\cdot\nG = u_{\nG}\ , \\
&m_{\tB} = (\mat{m}_\Gamma\cdot\nCo)\cdot\nCo &&\Longleftrightarrow &&\omega_{\tB} = \left[\mat{H}\cdot\vek{u} - \gradG{(\vek{u}\cdot\nG)}\right]\cdot\nCo\ , \\
& && &&\phantom{\omega_{\tB}}= -\left[(\gradGD{\vek{u})^\T\cdot\nG}\right]\cdot\nCo\ .
\end{align*}
In \autoref{tab:bc}, common support types are given. Other boundary conditions (e.g., membrane support,\,\ldots) may be derived, with the quantities above, accordingly.

\begin{table}[htb]
	\centering
	\begin{tabular}{l|c|c|c|c}
		Clamped edge & $u_{\tB} = 0$ & $u_{\nCo} = 0$ & $u_{\nG} = 0$ & $\omega_{\tB} = 0$\\ \hline
		Simply supported edge &$u_{\tB} = 0$ & $u_{\nCo} = 0$ & $u_{\nG} = 0$ & $m_{\tB} = 0$  \\ \hline		
		Symmetry support & $\tilde{p}_{\tB} = 0$ & $u_{\nCo} = 0$ & $\tilde{p}_{\nG} = 0$ & $\omega_{\tB} = 0$ \\ \hline
		Free edge & $\tilde{p}_{\tB} = 0$ & $\tilde{p}_{\nCo} = 0$ & $\tilde{p}_{\nG} = 0$ & $m_{\tB} = 0$	
	\end{tabular}
	\caption{Set of common boundary conditions}
	\label{tab:bc}
\end{table}

%% file: Content/Implementation.tex
\section{Implementational aspects}
\label{sec:impl}

The continuous weak form is discretized using isogeometric analysis as proposed by Hughes et al.~\cite{Hughes_2005a, Cottrell_2009a}. The NURBS patch $T$ is the middle surface of the shell and the elements $\tau_i\ ( i=1,\,\ldots,\,n_\t{Elem})$ are defined by the knot spans of the patch. The mesh is then defined by the union of the elements $\Gamma = \bigcup\limits_{\tau\in T} \tau\ .$\par

There is a fixed set of local basis functions $\lbrace N_i^k(\vek{r})\rbrace$ of order $k$ with $i = 1,\,\ldots,\,n_k$ being the number of control points and the displacements $\lbrace\hat{u}_i,\,\hat{v}_i,\,\hat{w}_i\rbrace$ stored at the control points $i$ are the degrees of freedom. Using the isoparametric concept, the shape functions $N_i^k(\vek{r})$ are NURBS of order $k$. The surface derivatives of the shape functions are computed as defined in \autoref{sec:pre} , similar as in the Surface FEM \cite{Dziuk_2013a, Demlow_2009a, Fries_2018b, Fries_2017b} using NURBS instead of Lagrange polynomials as ansatz and test functions. The shape functions of order $k \ge 2$  are in the function space $\mathcal{V}$, see \autoref{eq:funspaceV}. In fact, the used shape functions are in the Sobolev space $\mathcal{H}^k(\Gamma)^3 \subset \mathcal{V}$ iff $k \ge 2$. \par

The resulting element stiffness matrix $\mat{K}_{\t{Elem}}$ is a $3\times3$ block matrix and is divided into a membrane and bending part
\begin{align}
\mat{K}_{\t{Elem}} &= \mat{K}_{\t{Elem,M}} + \mat{K}_{\t{Elem,B}}\ .
\end{align}
The membrane part is defined by
\begin{align}
\mat{K}_{\t{Elem,M}} &=t\int_{\Gamma} P_{ib} \cdot [\hat{\mat{K}}]_{bj}\ \d \Gamma\\
[\hat{\mat{K}}]_{kj} &= \mu(\delta_{kj}\vek{N}_{,a}^{\Gamma}\cdot\vek{N}_{,a}^{\Gamma^\T} + \vek{N}_{,j}^{\Gamma}\cdot\vek{N}_{,k}^{\Gamma^\T}) + \lambda\vek{N}_{,k}^{\Gamma}\cdot\vek{N}_{,j}^{\Gamma^\T}\ ,
\end{align}
summation over $a$ and $b$. The matrix $\hat{\mat{K}}$ is determined by directional first-order derivatives of the shape functions $\vek{N}$. One may recognize that the structure of the matrix $\hat{\mat{K}}$ is similar to the stiffness matrix of $3$D linear elasticity problems. For the bending part we have
\begin{align}
[\mat{K}_{\t{Elem,B}}]_{ij} &= D_{\t{B}} \int_{\Gamma} n_in_j \tilde{\mat{K}}
\ \d \Gamma \\		
\tilde{\mat{K}} &= (1-\nu) \vek{N}^{\t{cov}}_{,ab}\cdot\vek{N}^{\t{cov}^\T}_{,ab} + \nu \vek{N}^{\t{cov}}_{,cc}\cdot\vek{N}^{\t{cov}^\T}_{,dd}\ .
\end{align}
A summation over $a,\,b$ on the one hand and $c,\,d$ on the other has to be performed. The first term of $\tilde{\mat{K}}$ is the contraction of the covariant Hessian matrix $\mat{H}\mat{e}^{\t{cov}}_\Gamma$ and the second term may be identified as the Bi-Laplace operator. Note that for the Bi-Laplace operator also directional derivatives may be used, due to the fact that the trace of second order derivatives is invariant, although the components differ. This suggests a further rearrangement of the contraction of the covariant Hessian matrix in order to use only directional derivatives, which is preferred from an implementational point of view. The equivalent expression of $\tilde{\mat{K}}$ using only second-order directional derivatives is
\begin{align}
\tilde{\mat{K}} &= (1-\nu) P_{ea}\vek{N}^{\t{dir}}_{,ab}\cdot\vek{N}^{\t{dir}^\T}_{,be} + \nu \vek{N}^{\t{dir}}_{,cc}\cdot\vek{N}^{\t{dir}^\T}_{,dd}\ ,
\end{align}
with summation over $a,\,b,\,e$ and $c,\,d$ as above. When the shell is given through a parametrization, the resulting element stiffness matrix in the classical theory, e.g., \cite{Cirak_2000a} is equivalent to the element stiffness matrix from above, but in the classical setting the computation \revStart may be found more \revEnd cumbersome due to fact that the local basis vectors and the metric tensor in co- and contra-variant form has to be computed. In contrast, herein, the surface gradients and second-order derivatives are first applied to the shape functions (NURBS or classical finite element functions) to obtain $\vek{N}_{,i}^\Gamma,\,\vek{N}_{,ij}^{\t{dir}}$ and $\vek{N}_{,ij}^{\t{cov}}$, which is independent of the application.\par
In this sense, a significant part of the complexity of implementing shells is shifted to finite element technology and may be recycled for any kind of boundary value problems on curved surfaces in $\mathbb{R}^3$. Examples are transport problems \cite{Dziuk_1988a, Demlow_2009a, Dziuk_2013a} or flow problems \cite{Fries_2018b, Jankuhn_2017a} on curved surfaces. We expect that future implementations in finite element software will provide frameworks for solving PDEs on manifolds and, based, e.g., on this work will also apply to shells.\revStart\ In order to emphasize the differences in the implementation, example Matlab\textregistered-codes for the proposed TDC-based formulation and the classical parametrization-based formulation are given in \autoref{sec:elemmat}, clearly highlighting the differences. \revEnd \par

The boundary conditions are weakly enforced by Lagrange multipliers \cite{Zienkiewicz_2013a}. The shape functions of the Lagrange multipliers are NURBS of the same order than the shape functions of the displacements. Therefore, the shape functions of the Lagrange multiplier $i$ is defined as
\begin{align}
\lbrace N^k_{i\,\t{L}}(\vek{r})\rbrace = \lbrace N_i^k(\vek{r})\rvert_{\p\Gamma_{\t{D}}}\rbrace\ .
\end{align}
For the test cases shown in \autoref{sec:numres}, bounded condition numbers and unique solutions are observed. The usual assembly yields a linear system of equations in the form
\begin{align}
\begin{bmatrix}
\mat{K} & \mat{C} \\
\mat{C}^\T & \mat{0}
\end{bmatrix} \cdot \begin{bmatrix}
\hat{\underline{\vek{u}}}\\
\hat{\vek{\lambda}}
\end{bmatrix} = \begin{bmatrix}
\vek{f} \\
\vek{0}
\end{bmatrix}\ ,
\end{align}  
with $[\hat{\underline{\vek{u}}},\, \hat{\vek{\lambda}}]^\T = [\hat{\vek{u}},\,\hat{\vek{v}},\,\hat{\vek{w}},\,\hat{\vek{\lambda}}]$ being the sought displacements of the control points and Lagrange multipliers. With the shape functions of the Lagrange multipliers $\vek{N}_\t{L}$, the constraint matrix $\mat{C}$ for simply supported edges is defined by
\begin{align}
\mat{C}^\T = \int_{\p\Gamma_{\t{D}}}\begin{bmatrix}
\vek{N}_{\t{L}} \cdot \vek{N}^\T & \mat{0} & \mat{0} \\
\mat{0}& \vek{N}_{\t{L}} \cdot \vek{N}^\T & \mat{0} \\
\mat{0} & \mat{0} & \vek{N}_{\t{L}} \cdot \vek{N}^\T
\end{bmatrix} \d s\ ,
\end{align}
for clamped edges
\begin{align}
\mat{C}^\T = \int_{\p\Gamma_{\t{D}}}\begin{bmatrix}
\vek{N}_{\t{L}} \cdot \vek{N}^\T & \mat{0} & \mat{0} \\
\mat{0}& \vek{N}_{\t{L}} \cdot \vek{N}^\T & \mat{0} \\
\mat{0} & \mat{0} & \vek{N}_{\t{L}} \cdot \vek{N}^\T \\
\vek{N}_{\t{L}} \cdot (n_x {n}_{\p\Gamma_i} \vek{N}_{,i}^{{\Gamma}^\T}) & \vek{N}_{\t{L}} \cdot (n_y {n}_{\p\Gamma_i} \vek{N}_{,i}^{{\Gamma}^\T}) & \vek{N}_{\t{L}} \cdot (n_z {n}_{\p\Gamma_i} \vek{N}_{,i}^{{\Gamma}^\T})
\end{bmatrix} \d s \ ,
\end{align}
and for symmetry supports
\begin{align}
\mat{C}^\T = \int_{\p\Gamma_{\t{D}}}\begin{bmatrix}
{n}_{\p\Gamma_x}\vek{N}_{\t{L}} \cdot \vek{N}^\T & {n}_{\p\Gamma_y}\vek{N}_{\t{L}} \cdot \vek{N}^\T & {n}_{\p\Gamma_z}\vek{N}_{\t{L}} \cdot \vek{N}^\T \\
\vek{N}_{\t{L}} \cdot (n_x {n}_{\p\Gamma_i} \vek{N}_{,i}^{{\Gamma}^\T}) & \vek{N}_{\t{L}} \cdot (n_y {n}_{\p\Gamma_i} \vek{N}_{,i}^{{\Gamma}^\T}) & \vek{N}_{\t{L}} \cdot (n_z {n}_{\p\Gamma_i} \vek{N}_{,i}^{{\Gamma}^\T})
\end{bmatrix} \d s \ .
\end{align}
Note that all constraint matrices have three block-columns refering to the unknowns $\hat{\vek{u}}, \hat{\vek{v}}, \hat{\vek{w}}$.

%% file: Content/NumRes.tex
\section{Numerical results}
\label{sec:numres}

The numerical results are achieved using NURBS functions for the geometry and shape function definition, following the methodology of isogeometric analysis (IGA) \cite{Hughes_2005a, Kiendl_2009a, Benson_2010a, Nguyen_2015a, Guo_2017a}. The definition of NURBS is omitted here for brevity but is found at numerous references in the frame of IGA, e.g., \cite{Piegl_1997a, Cottrell_2009a}.

The obtained shell equations are carefully verified and compared to the classical approach with different test cases. As already mentioned above the proposed approach leads to an equivalent stiffness matrix for arbitrary curved and non-curved shells. Consequently, the same convergence properties as shown, e.g., in \cite{Kiendl_2009a, Cirak_2000a} are expected. In the following, the results of the convergence analyses of a flat shell embedded in $\mathbb{R}^3$, the Scordelis-Lo roof, and the pinched cylinder test (part of the shell obstacle course proposed by Belytschko et al.~\cite{Belytschko_1985a}) are shown. Furthermore, a new test case with a challenging geometry is proposed which features smooth solutions enabling higher-order convergence rates. These rates are confirmed in the residual error as no analytic solution exists, see \autoref{sec:ex4}. Other examples (e.g., pinched hemispherical shell, shells of revolution, etc.) have been considered but are omitted here for brevity.\par

In the convergence studies, NURBS patches with different orders and numbers of knot spans in each direction are employed. This is equivalent to meshes with higher-order elements and $n=\lbrace2,\,4,\,8,\,16,\,32\rbrace$ elements per side are used. The orders are varied as $p=\lbrace2,\,3,\,4,\,5,\,6\rbrace$.

\subsection{Flat shell embedded in $\mathbb{R}^3$}
\label{sec:ex1}

\begin{figure}[htb]
	\begin{minipage}[h]{.5\textwidth}\centering
		\includegraphics[trim={2.5cm 0 2.5cm 0},clip,width=.95\textwidth, height = .9\textwidth, keepaspectratio]{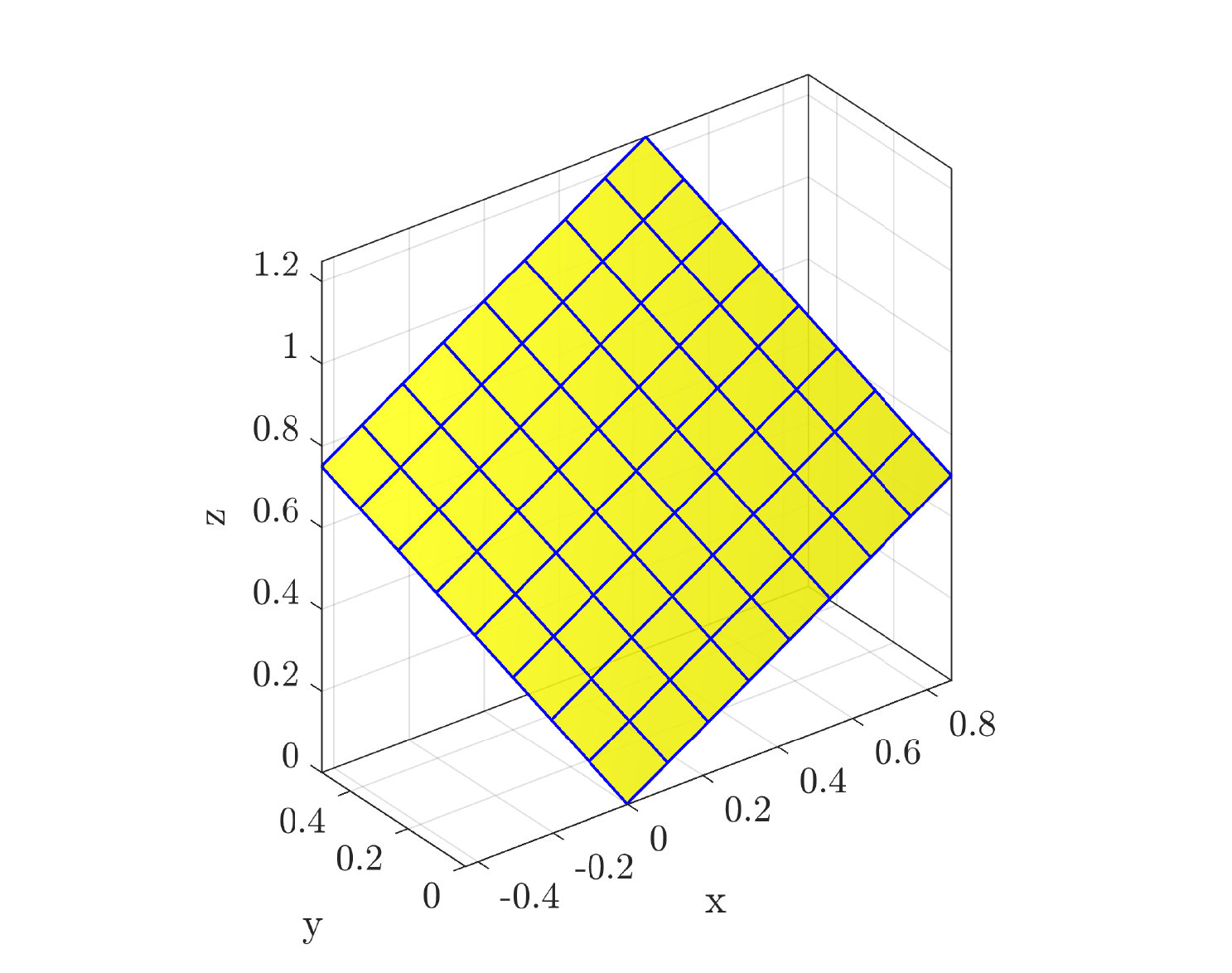}
	\end{minipage}
	\begin{minipage}[h]{.48\textwidth}\scriptsize\flushleft
		\begin{tabular}[h]{ll}
			Geometry: & Quadrilateral flat shell\\
			& $L = 1$ \\
			& $t = \SI{0.01}{}$\\
			& $\vek{n}_\Gamma~=~[-\sfrac{1}{4},\,-\sfrac{\sqrt{3}}{2},\,\sfrac{\sqrt{3}}{4}]^\T$\\[.25 cm]
			Material parameters: & $E = \SI{10000}{}$ \\
			& $\nu = \SI{0.3}{}$\\[.25 cm]
			Load: & $\vek{f}_t $ and $f_n$ \\[.25 cm]
			Support: & Simple support on all edges
		\end{tabular}
	\end{minipage}
	\caption{Definition of flat shell problem.}
	\label{fig:overplate}
\end{figure}

Following a similar rationale as in \cite{Hansbo_2017a}, as a first test case, we consider a simple quadrilateral, flat shell with the normal vector $\vek{n}_\Gamma = [-\sfrac{1}{4},\,-\sfrac{\sqrt{3}}{2},\,\sfrac{\sqrt{3}}{4}]^\T$ in $\mathbb{R}^3$, see \autoref{fig:overplate}. The shell is simply supported at all edges. For verification, the load vector $\vek{f}$ is split into tangential $\vek{f}_t$ and normal $f_n$ loads. The tangential loads are obtained with the method of manufactured solution for a given displacement field $\vek{u}_t(\vek{x}) = \left[\, [1, 1]^\T \cdot \sfrac{1}{4} \cdot \sin(\pi r)\sin(\pi s) \right] \circ \vek{\chi}^{-1}$. In normal direction, a sinusoidal load $f_n(\vek{x}) = \left[-D \sin(\pi r)\sin(\pi s)\right]\circ \vek{\chi}^{-1}$ is applied to the shell. Herein, $\vek{\chi}$ is an affine mapping function (rigid-body rotation) from the horizontal parameter space to the real domain. An analytic solution for the normal displacements is easily obtained with $u_n(\vek{x}) = \left[-(\sin(\pi r)\sin(\pi s))/(4\pi^4)\right] \circ \vek{\chi}^{-1}$, \cite{Timoshenko_1959a}.  The shell is defined with $L = \SI{1}{}$ and the thickness is set to $t = \SI{0.01}{}$. The material parameters are: Young's modulus $E = \SI{10000}{}$ and the Poisson's ratio $\nu = \SI{0.3}{}$.

\begin{figure}[htb]
	\centering
	\subfloat[front view]{\includegraphics[width=.48\textwidth, height = .42\textwidth, keepaspectratio]{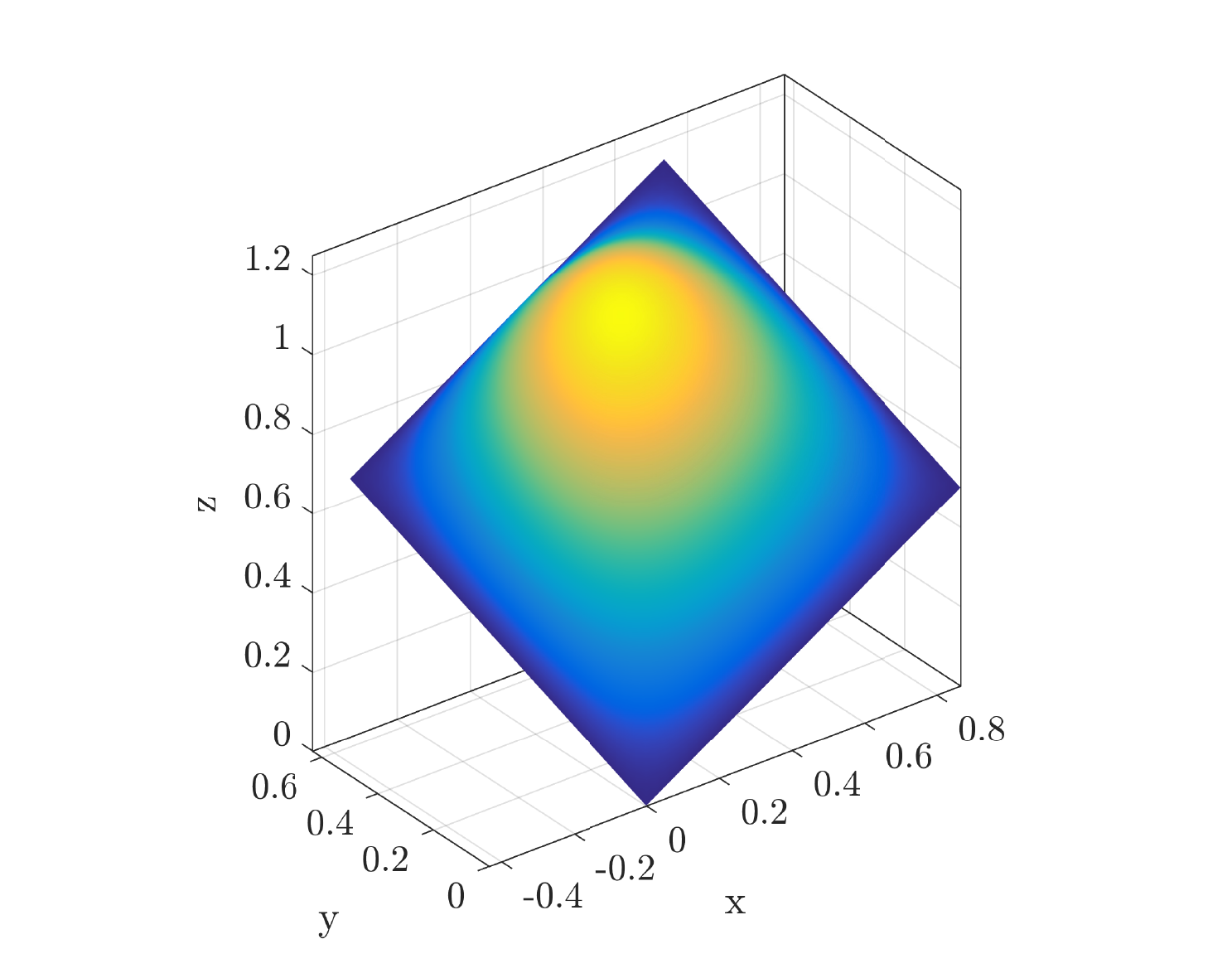}}
	\hfil
	\subfloat[rotated view]{\includegraphics[trim={3.cm 0 0 0},clip,width=.48\textwidth, height = .42\textwidth, keepaspectratio]{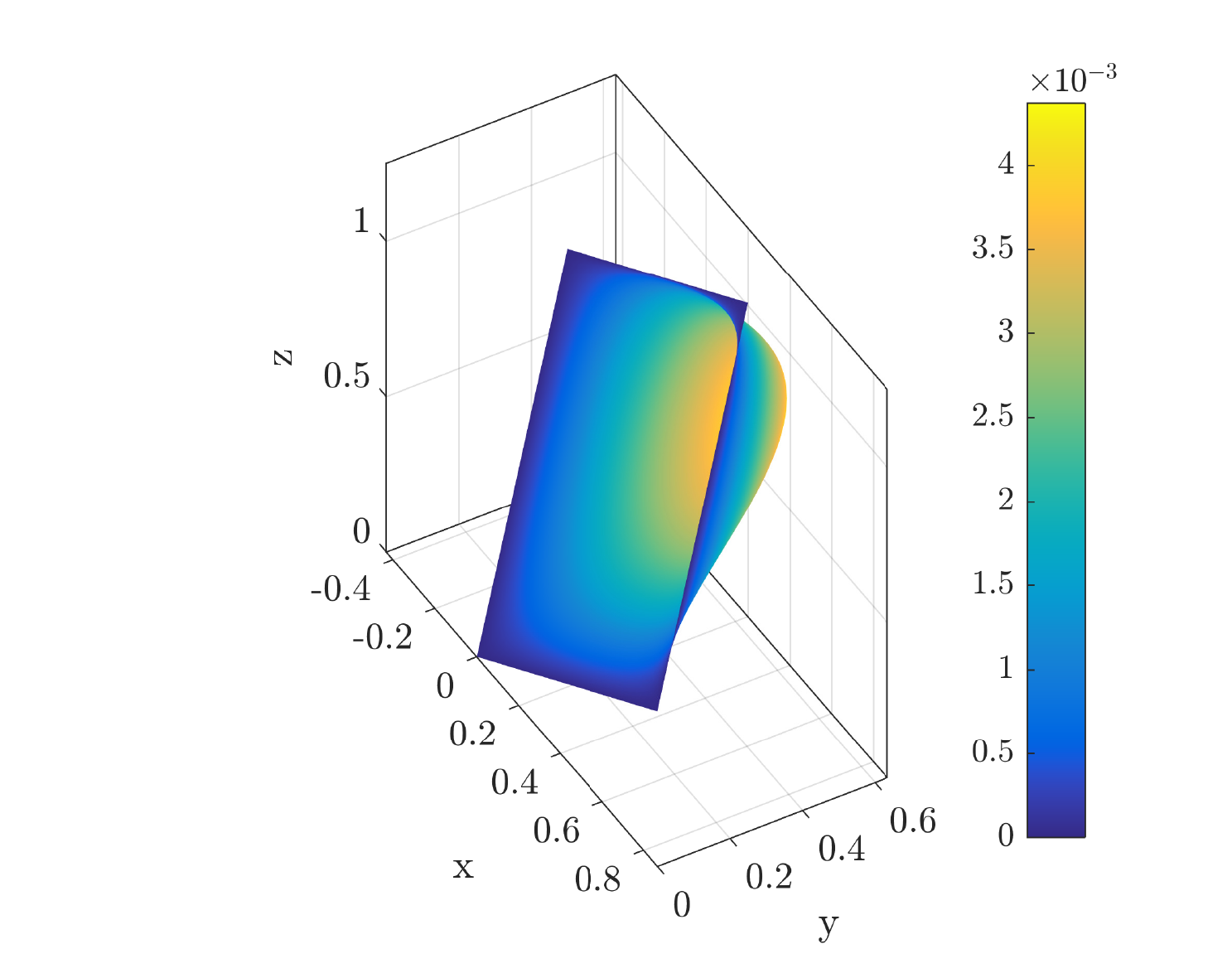}}
	\caption{Displacement $\vek{u}$ of arbitrarily orientated flat shell, scaled by two orders of magnitude.}
	\label{fig:platedisp}
\end{figure}

In \autoref{fig:platedisp}, the solution of the shell is illustrated. The displacements are scaled by two orders of magnitude. The colours on the deformed surface indicate the Euclidean norm of the displacement field $\Vert \vek{u} \Vert$.

\begin{figure}[h]
	\centering
	\subfloat[Relative $L_2$-norm of displacements $\vek{u}$]{\includegraphics[width=.45\textwidth, height = .45\textwidth, keepaspectratio]{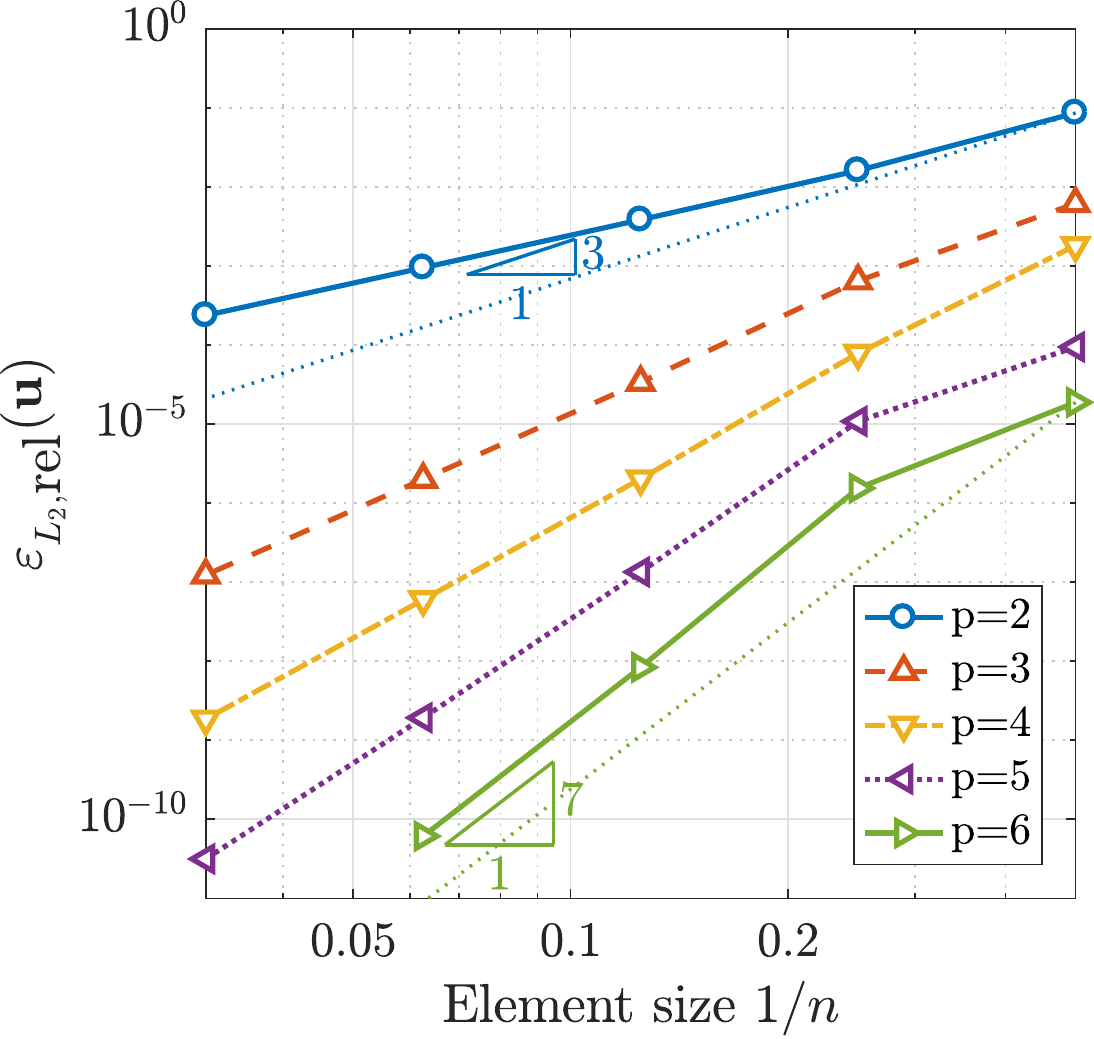}\label{fig:platel2u}}
	\hfil
	\subfloat[Relative $L_2$-norm of normal forces $\mat{n}^{\t{real}}_\Gamma$]{\includegraphics[width=.45\textwidth, height = .45\textwidth, keepaspectratio]{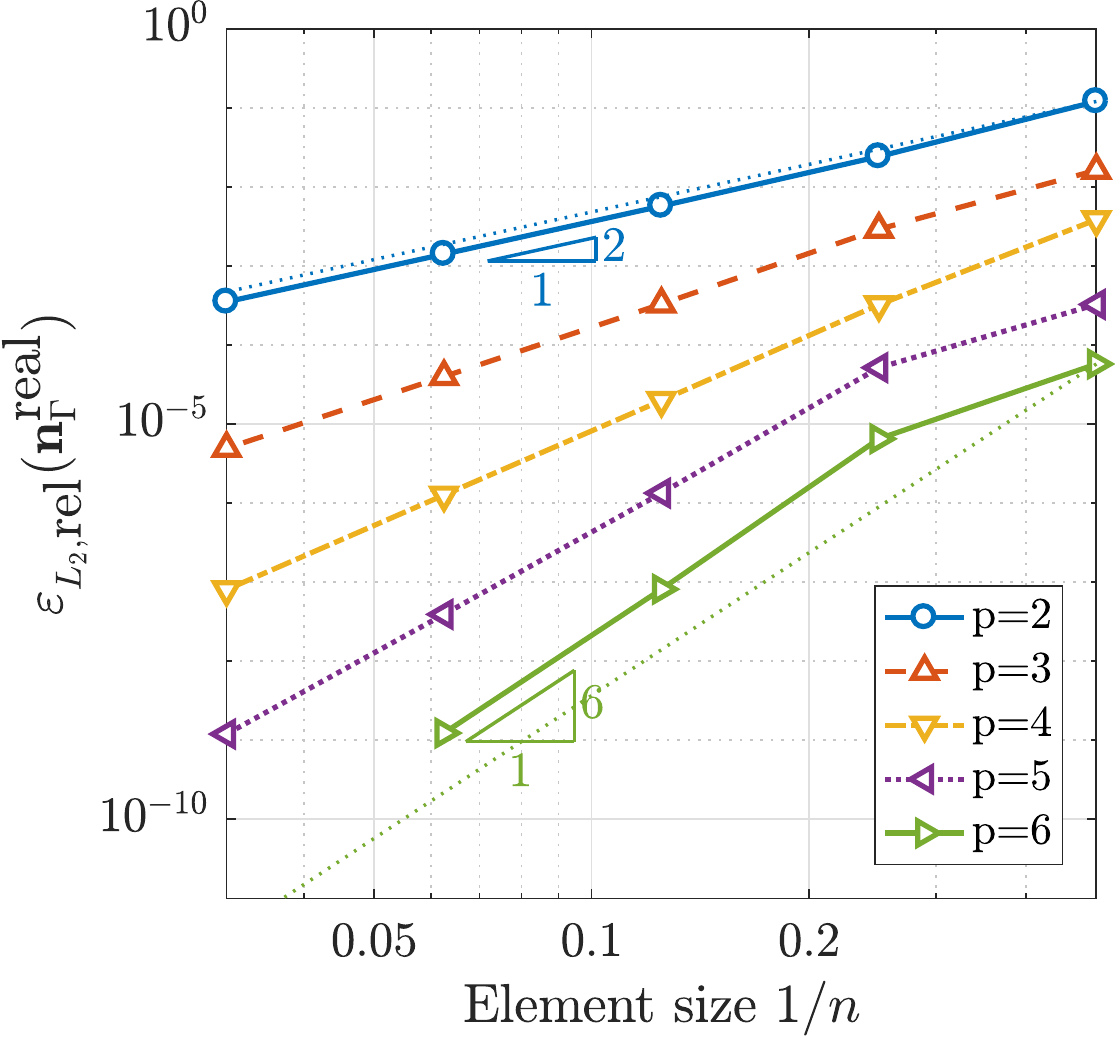}\label{fig:platel2n}}\\
	\subfloat[Relative $L_2$-norm of bending moments $\mat{m}_{\Gamma}$]{\includegraphics[width=.45\textwidth, height = .45\textwidth, keepaspectratio]{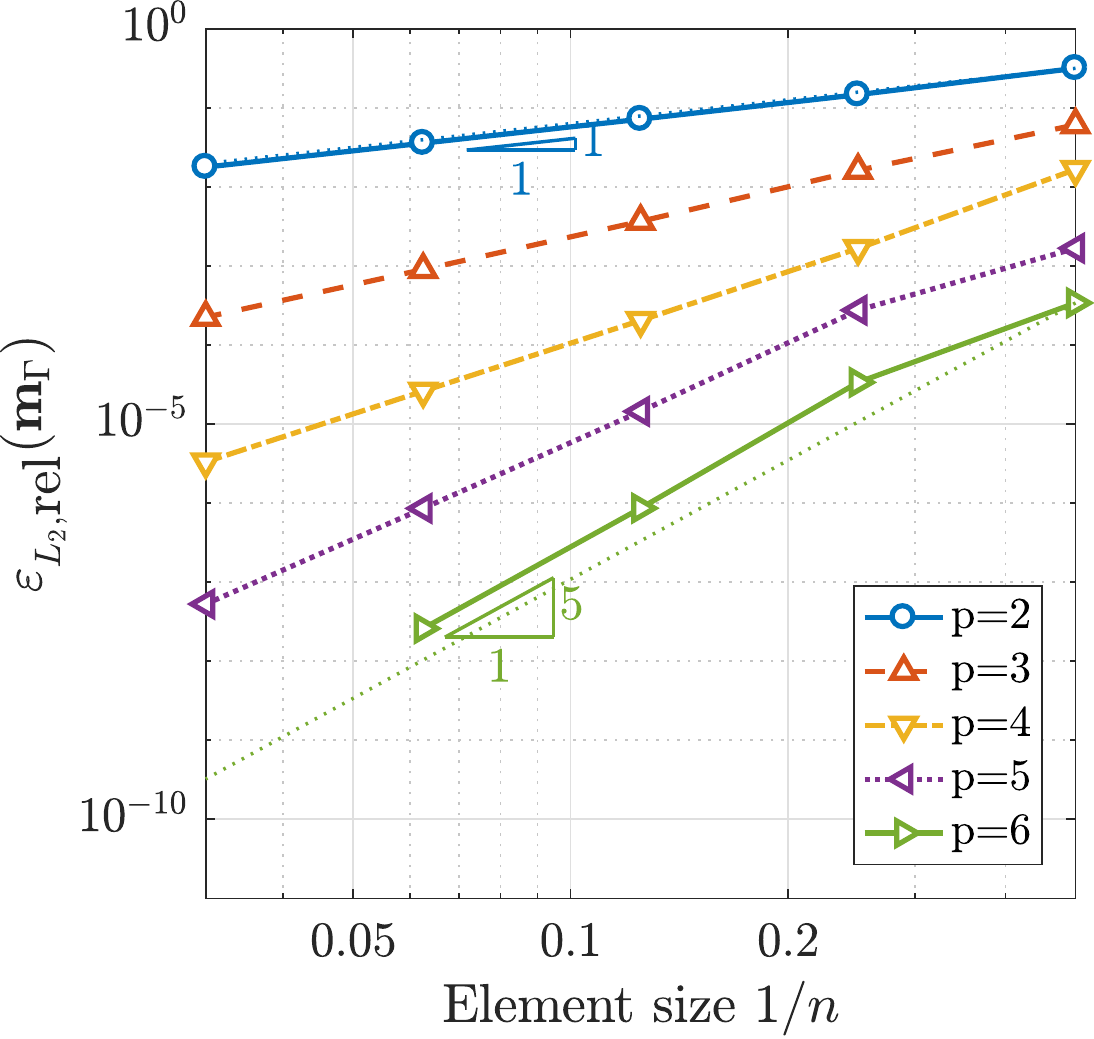}\label{fig:platel2m}}
	\hfil 	
	\subfloat[Relative $L_2$-norm of transverse shear forces~$\vek{q}$]{\includegraphics[width=.45\textwidth, height = .45\textwidth, keepaspectratio]{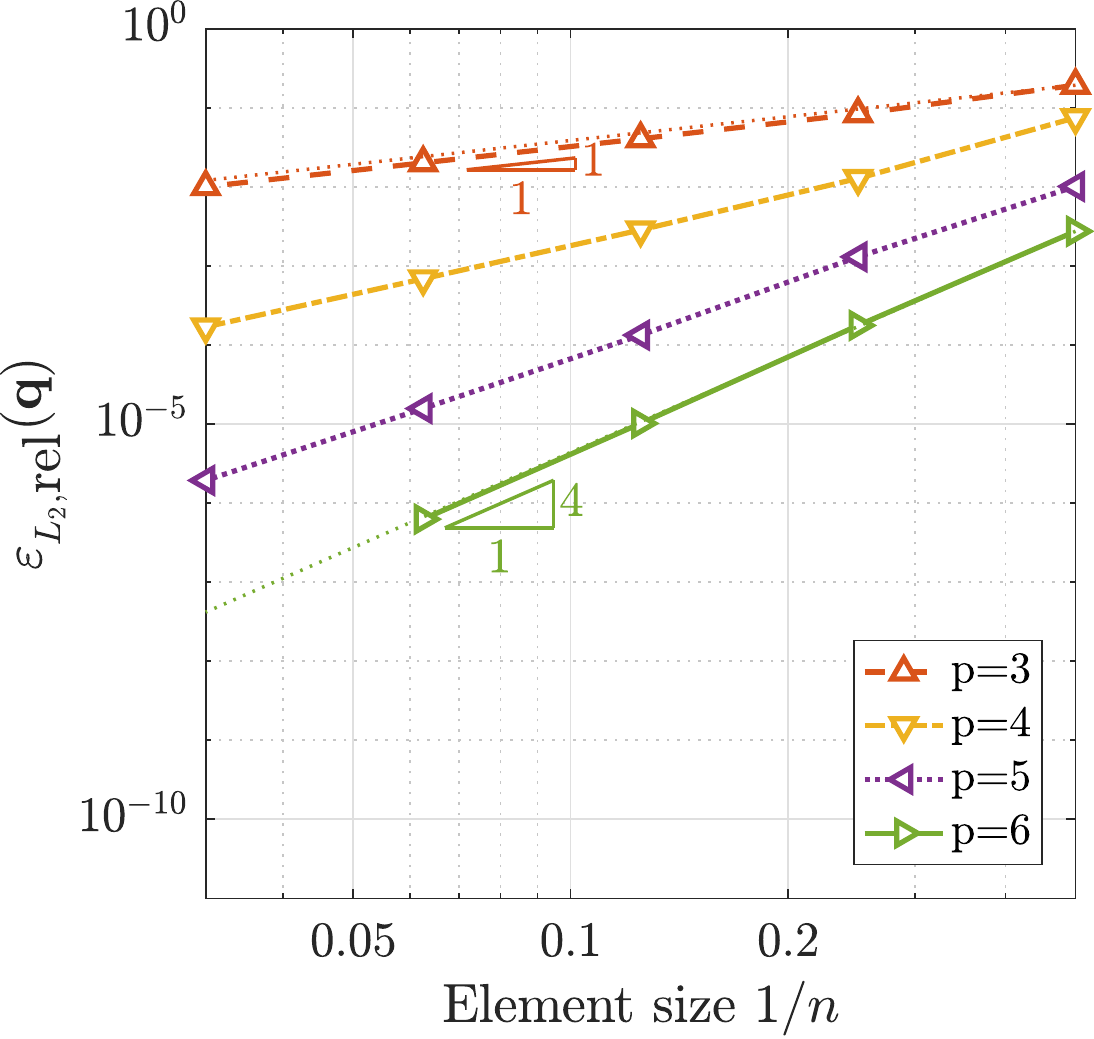}\label{fig:platel2q}}
	\caption{Convergence results for the rotated flat shell.}
	\label{fig:platel2}
\end{figure}

The results of the convergence analysis are shown in \autoref{fig:platel2}. The curves are plotted as a function of the element size $\sfrac{1}{n}$ (which is rather a characteristic length of the knot spans). The dotted lines indicate the theoretical optimal order of convergence.
In \autoref{fig:platel2u}, the relative $L_2$-error of the primal variable (displacements) is shown. Optimal higher-order convergence rates $\mathcal{O}(p+1)$ are achieved. In the figures \autoref{fig:platel2n} to \autoref{fig:platel2q}, the relative $L_2$-errors of the normal forces (membrane forces), bending moments and transverse shear forces are plotted. For all stress resultants the theoretical optimal orders of convergence are achieved. It is clear that the same results were obtained if the results are computed for the purely two-dimensional case as, e.g., in \cite{Cirak_2000a}.

\clearpage
\subsection{Scordelis-Lo roof}
\label{sec:ex2}

\begin{figure}[ht]
	\begin{minipage}{.5\textwidth}\centering
		\includegraphics[width=.95\textwidth, height = .9\textwidth, keepaspectratio]{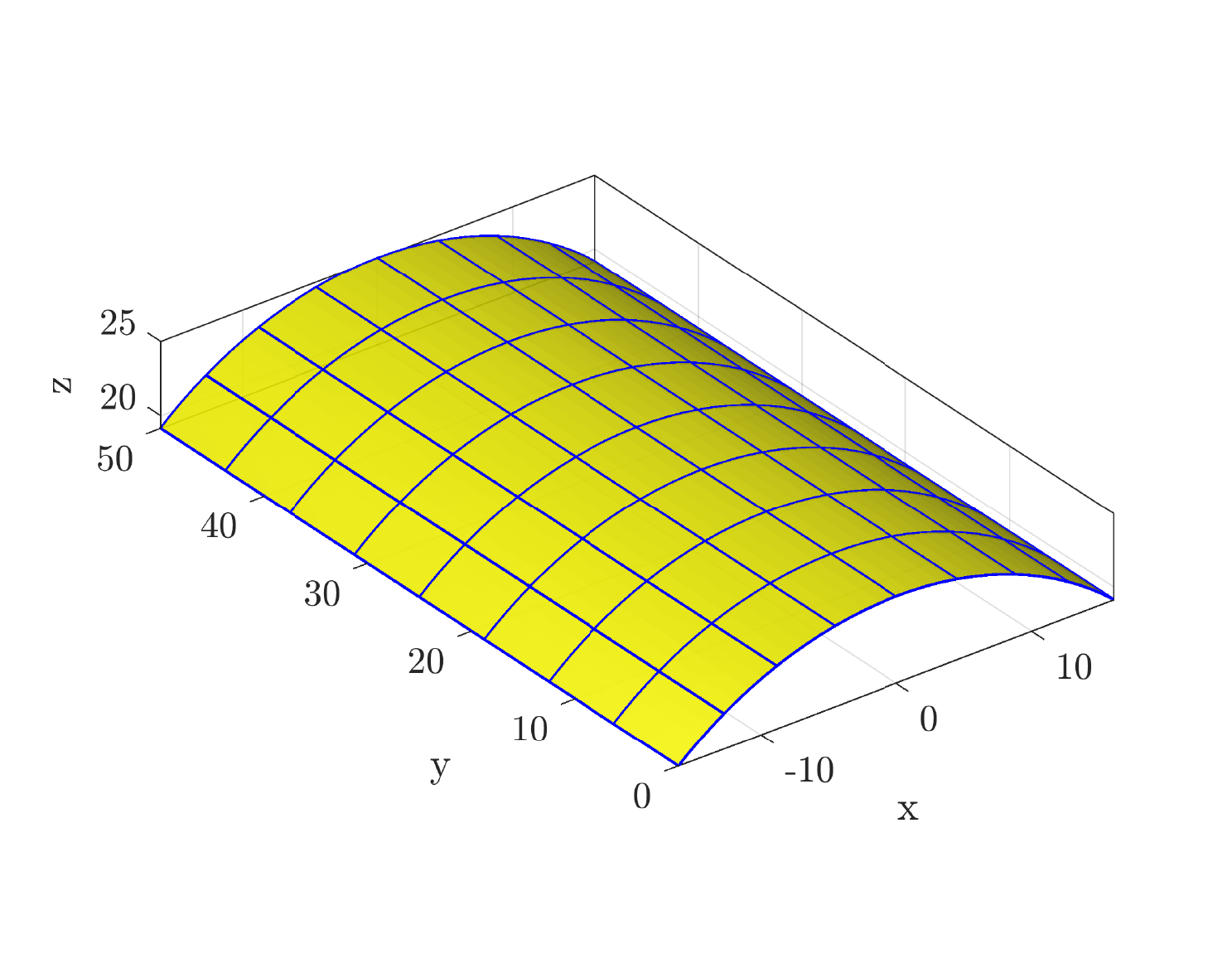}
	\end{minipage}
	\begin{minipage}{.48\textwidth}\scriptsize\flushleft
		\begin{tabular}[h]{ll}
			Geometry: & Cylindrical shell\\
			& $L = 50$ \\
			& $R = 25 $ \\
			& $\phi = \SI{80}{\degree}$ \\
			& $t = \SI{0.25}{}$\\[.25 cm]
			Material parameters: & $E = \SI{4.32e8}{}$ \\
			& $\nu = \SI{0.0}{}$\\[.25 cm]
			Load: & Gravity load $\vek{f} = [0,\,0,\,-90]^\T $\\[.25 cm]
			Support: & Rigid diaphragms at it ends
		\end{tabular}
	\end{minipage}
	\caption{Definition of Scordelis-Lo roof problem.}
	\label{fig:overscor}
\end{figure}

The Scordelis-Lo roof is a cylindrical shell and is supported with two rigid diaphragms at the ends. The shell is loaded by gravity forces, see \autoref{fig:overscor}. 
The cylinder is defined with $L = \SI{50}{},\ R = \SI{25}{}$ and the angle subtended by the roof is $\phi = \SI{80}{\degree}$. The thickness of the shell is set to $t = \SI{0.25}{}$. The material parameters are: Young's modulus $E = \SI{4.32e8}{}$ and the Poisson's ratio $\nu = \SI{0.0}{}$. In contrast to the first example, the maximum vertical displacement $u_{z,\max}$ is compared with the reference solution $u_{z,\max,\text{Ref}} = \SI{0.3024}{}$ as given in reference \cite{Belytschko_1985a}.

\begin{figure}[ht]
	\centering
	\subfloat[displacement $\vek{u}$]{\includegraphics[width=.52\textwidth, height = .52\textwidth, keepaspectratio]{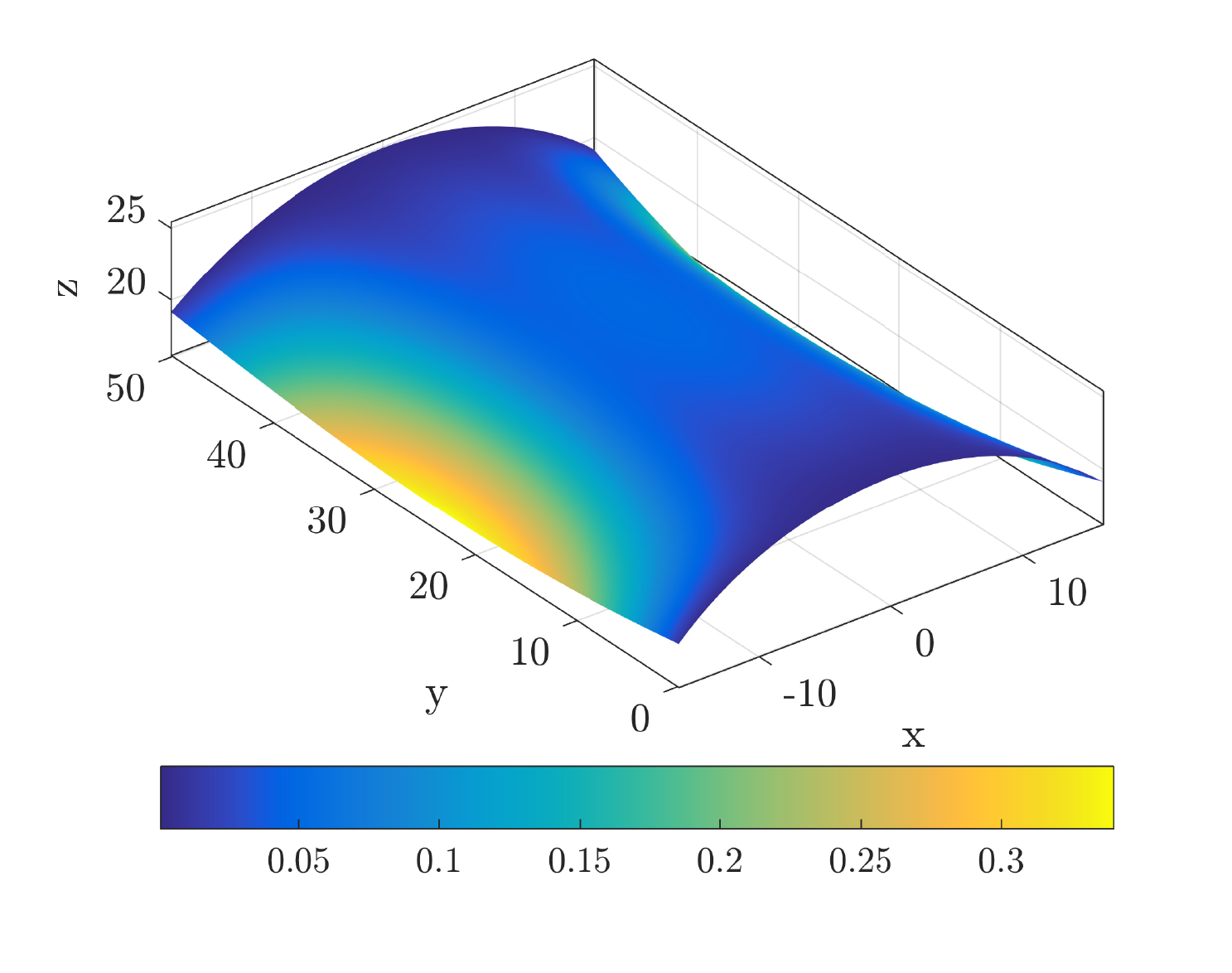}\label{fig:scordelisdisp}}
	\hfil
	\subfloat[convergence]{\includegraphics[width=.45\textwidth, height = .45\textwidth, keepaspectratio]{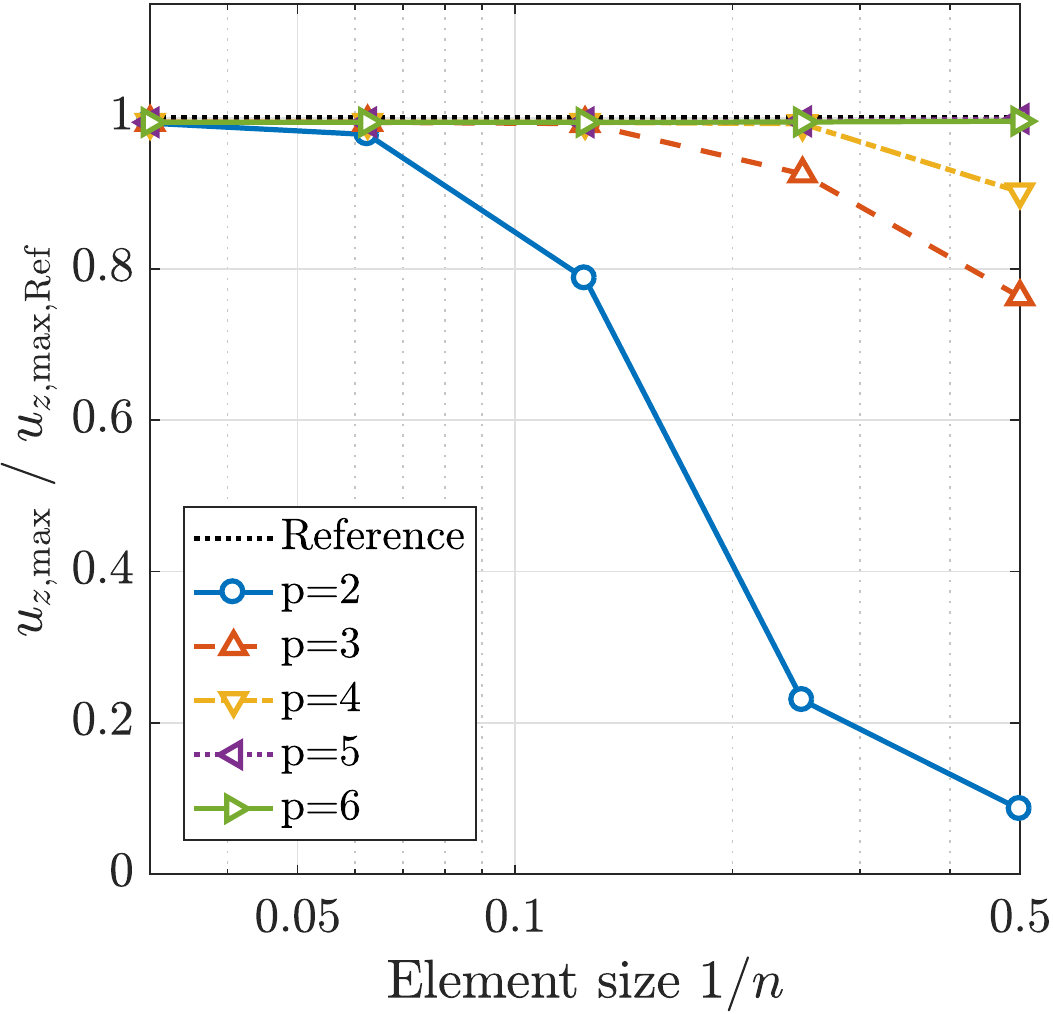}\label{fig:scordelisl2}}
	\caption{(a) Displacement field of the Lo-Scordelis roof scaled by one order of magnitude, (b) normalized convergence of reference displacement $u_{z,\max,\text{Ref}} = \SI{0.3024}{}$.}
\end{figure}

In \autoref{fig:scordelisdisp}, the numerical solution of the Scordelis-Lo roof is illustrated. The displacements are magnified by one order of magnitude.

In \autoref{fig:scordelisl2}, the convergence of the maximum displacement $u_{z,\max}$ is plotted up to polynomial order of $p=6$ as a function of the element (knot span) size. It is clearly seen that the expected results are achieved, with increasing accuracy for higher-order NURBS. Due to the lack of a more accurate reference solutions, it is not useful to show these results in a double-logarithmic diagram as usual for error plots. The style of presentation follows those of many other references such as, e.g., in \cite{Belytschko_1985a, Cirak_2000a, Kiendl_2009a}.

\subsection{Pinched cylinder}
\label{sec:ex3}

\begin{figure}[h]
	\begin{minipage}{.5\textwidth}\centering
		\includegraphics[width=.95\textwidth, height = .9\textwidth, keepaspectratio]{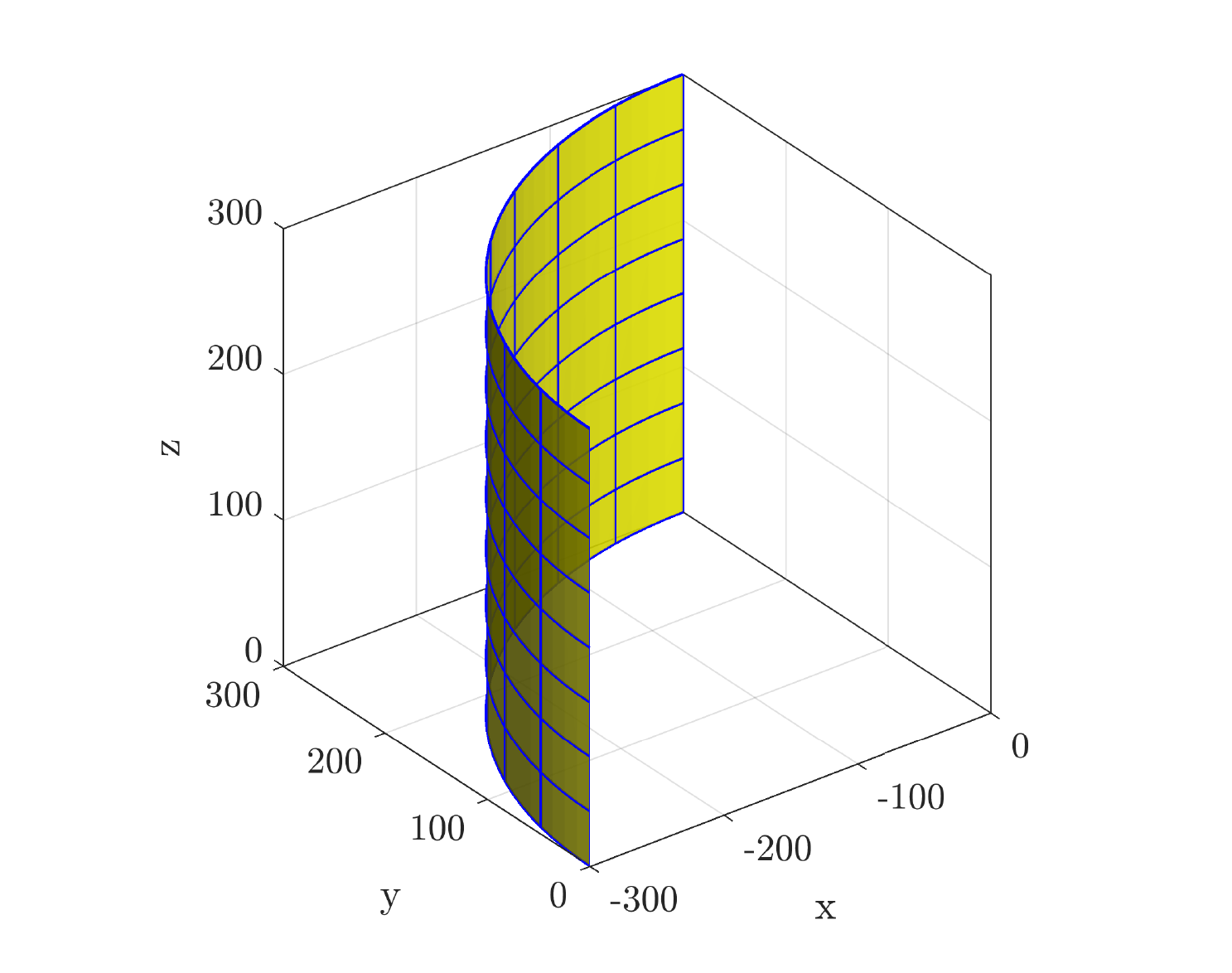}
	\end{minipage}
	\begin{minipage}{.48\textwidth}\scriptsize\flushleft
		\begin{tabular}[h]{ll}
			Geometry: & Cylinder\\
			&(one eighth of cylinder modeled)\\[.1 cm]
			& $\sfrac{L}{2} = 300$ \\
			& $R = 300 $ \\
			& $\phi = \SI{90}{\degree}$ \\
			& $t = \SI{3}{}$\\[.25 cm]
			Material parameters: & $E = \SI{3e6}{}$ \\
			& $\nu = \SI{0.3}{}$\\[.25 cm]
			Load: & Single unit forces\\[.25 cm]
			Support: & \parbox{5cm}{Rigid diaphragms at the top and\\ symmetry boundary conditions}
		\end{tabular}
	\end{minipage}
	\caption{Definition of the pinched cylinder problem.}
	\label{fig:overpcyl}
\end{figure}

The next test case is a cylindrical shell pinched with two diametrically opposite unit loads located within the middle of the shell, see \autoref{fig:overpcyl}. The cylinder is defined with $ L = \SI{600}{},\ R = \SI{300}{}$. The thickness is set to $t = \SI{3}{}$. The material properties are: Young's modulus $E = \SI{3e6}{}$ and the Poisson's ratio  $\nu = \SI{0.3}{}$. The  reference displacement at the loading points are $u_{\text{Ref}} = \SI{1.82488E-5}{}$ as given in reference \cite{Belytschko_1985a}. Due to symmetry only one eighth of the geometry is modelled.
 
\begin{figure}[h]
	\centering
	\subfloat[displacements $\vek{u}$]{\includegraphics[width=.52\textwidth, height = .52\textwidth, keepaspectratio]{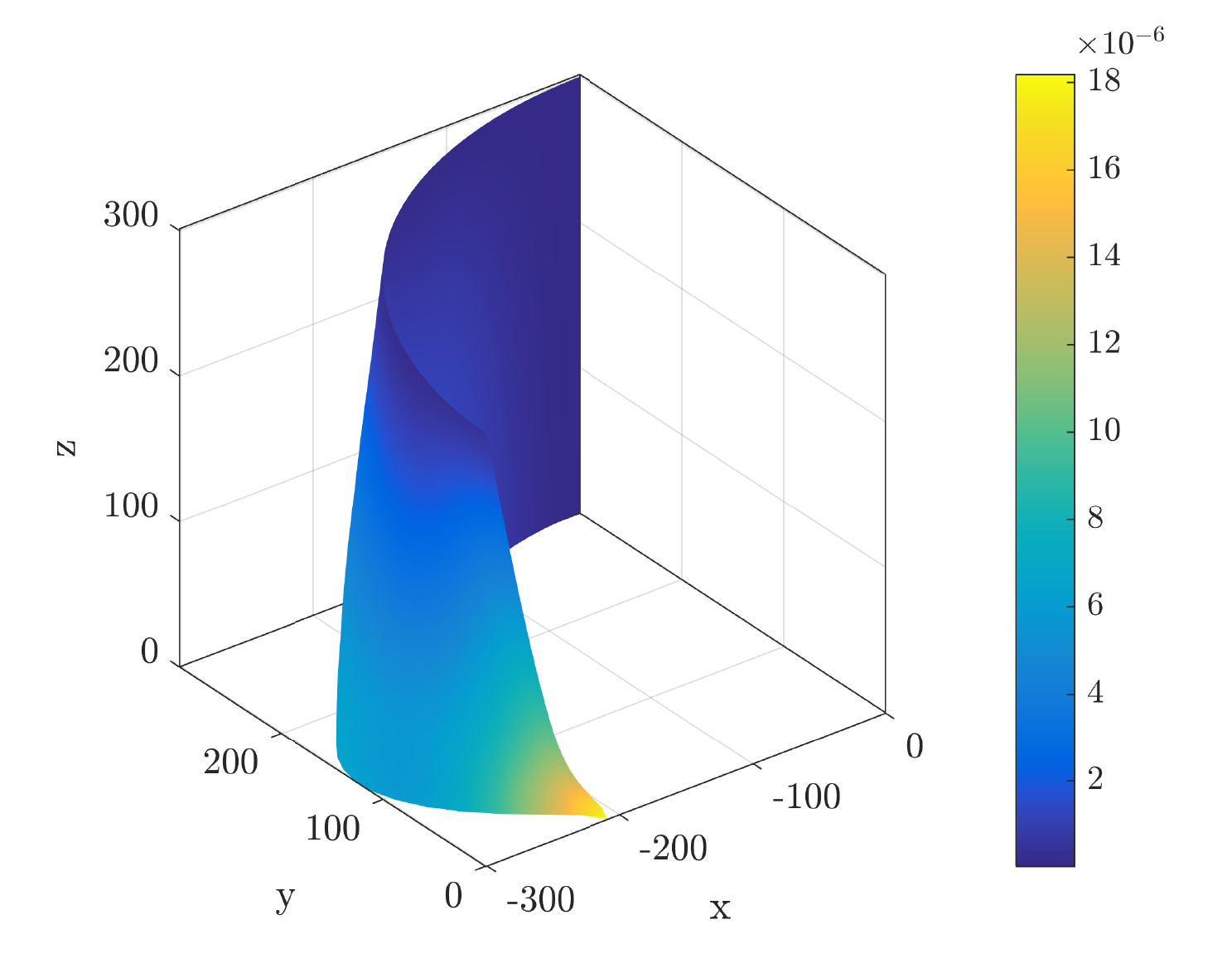}\label{fig:pcylinderdisp}}
	\hfil
	\subfloat[convergence]{\includegraphics[width=.45\textwidth, height = .45\textwidth, keepaspectratio]{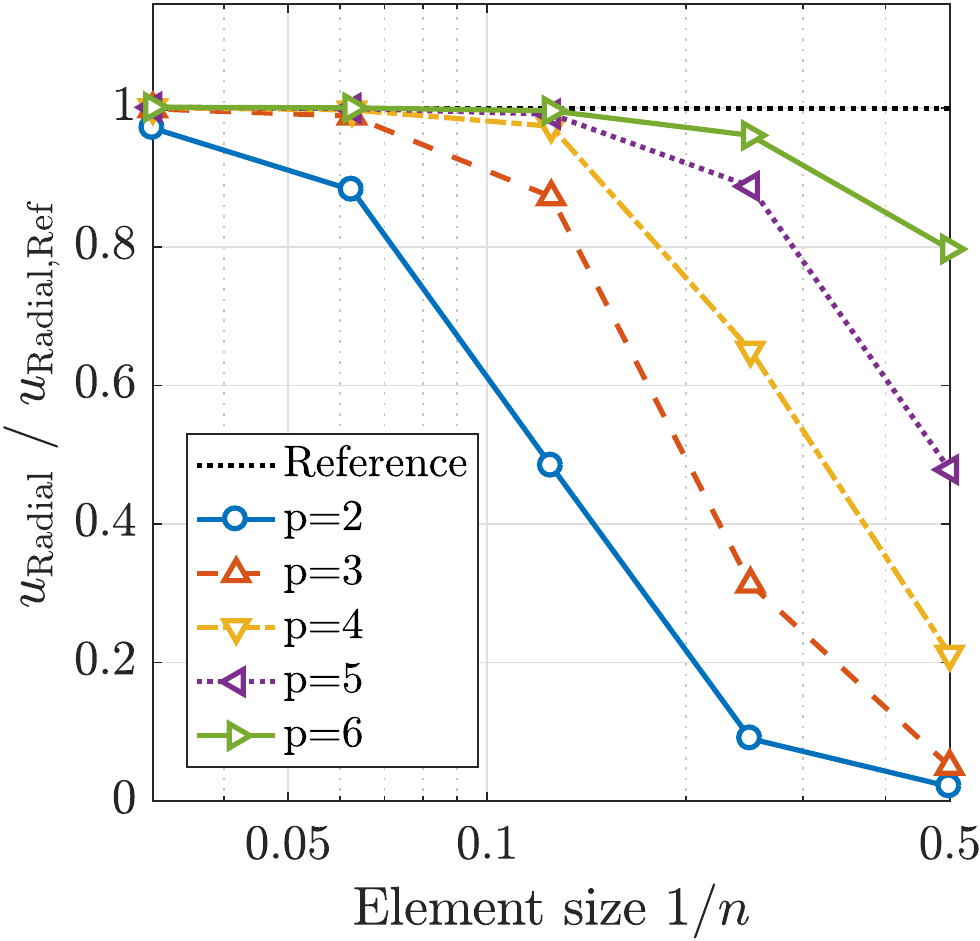}\label{fig:pcylinderl2}}
	\caption{Pinched cylinder: (a) Displacement $\vek{u}$ of one eighth of the geometry (scaled by  a factor of $\SI{5E6}{}$), (b) Normalized convergence of reference dis-
		placement $u_{\text{Radial},\text{Ref}} = \SI{1.82488E-5}{}$ at loading points.}
\end{figure}

In \autoref{fig:pcylinderdisp}, the numerical solution of the pinched cylinder is illustrated with scaled displacements by a factor of $\SI{5E6}{}$.

As in the example before, in \autoref{fig:pcylinderl2}, the convergence to a normalized reference displacement as a function of the element size is plotted. The results confirm with the expected convergence behaviour as shown in \cite{Cirak_2000a, Kiendl_2009a}. It is noted that due to the singularity in some mechanical quantities due to the single force, higher-order convergence rates are not possible here. However, the improvement for increasing the order of the NURBS is still seen in the figure. An additional grading of the elements in order to better resolve the singularity would have further improved the situation but is omitted here. 

\subsection{Flower shaped shell}
\label{sec:ex4}

\begin{figure}[h]
	\begin{minipage}{.5\textwidth}\centering
		\includegraphics[width=.9\textwidth, height = .9\textwidth, keepaspectratio]{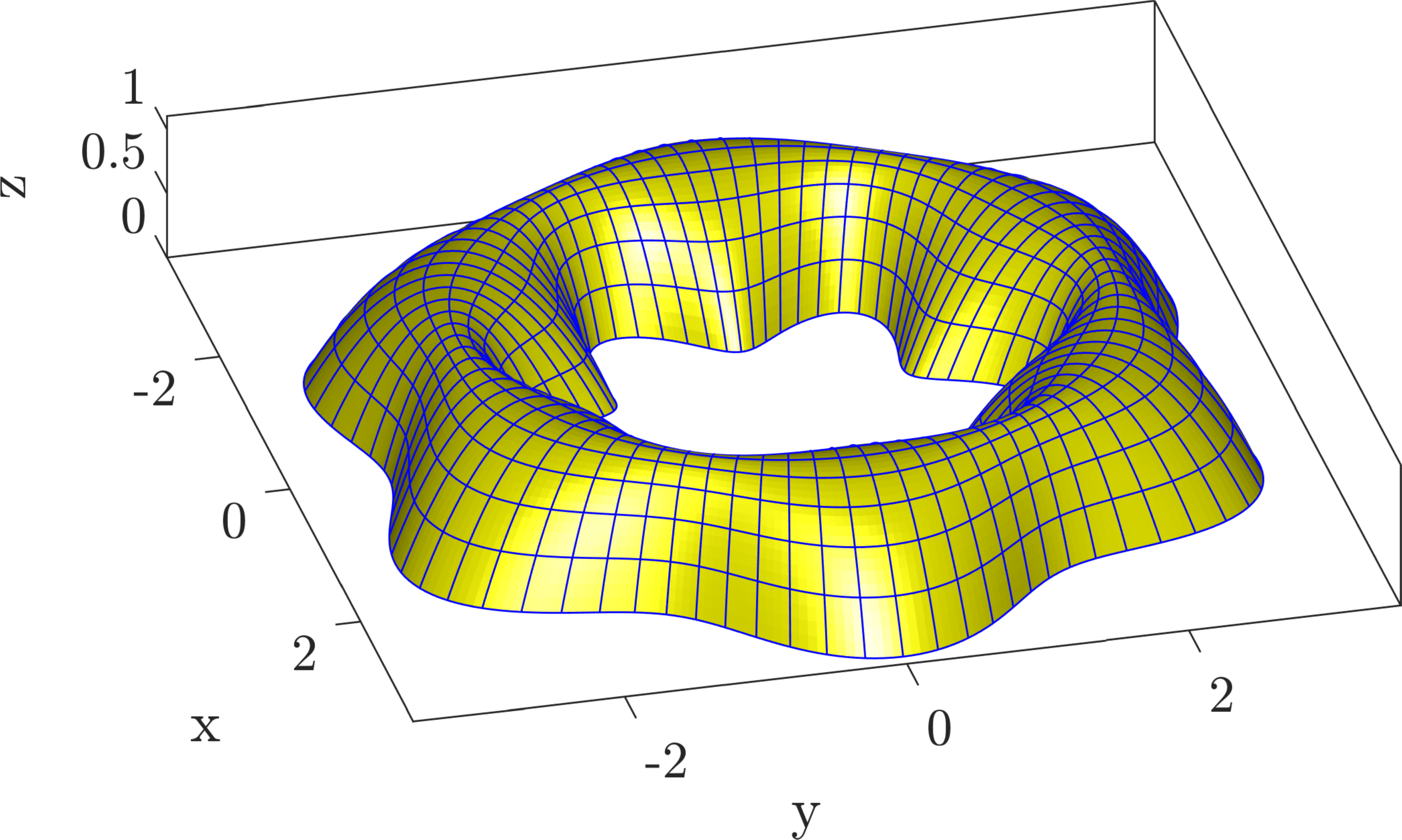}
	\end{minipage}
	\begin{minipage}{.48\textwidth}\scriptsize\flushleft
		\begin{tabular}[h]{ll}
			Geometry: & Flower shaped shell\\
			& see \autoref{eq:ex4geom} \\
			& $t = \SI{0.1}{\metre}$\\[.25 cm]
			Material parameters: & $E = \SI{1e5}{\kilo\newton\per\metre\squared}$ \\
			& $\nu = \SI{0.3}{}$\\[.25 cm]
			Load: & $\vek{f}(\vek{x}_\Gamma) = [1,\,2,\,-10]^\T \SI{}{\kilo\newton\per\metre\squared}$\\[.25 cm]
			Support: & Clamped edges at inner and \\
			&outer boundary
		\end{tabular}
	\end{minipage}
	\caption{Definition of flower shaped shell problem}
	\label{fig:overflower}
\end{figure}

As a last example, a more complex geometry is considered, which enables smooth mechanical fields and thereby enables higher-order convergence rates. The geometry of the middle surface is given with\\
\begin{minipage}{.4\textwidth}
	\begin{align*}
	\vek{x}_\Gamma(r,\,s) &= \begin{bmatrix}
	(A - C) \cos(\theta)\\
	(A - C) \sin(\theta)\\
	1-s^2
	\end{bmatrix}
	\end{align*}
\end{minipage}
\hfil
\begin{minipage}{.59\textwidth}
\begin{align}
\begin{split}
\t{with:}\quad
&r,\,s \in [-1,\,1]\ ,\ A = 2.3\ ,\ B = 0.8 \\
&\theta(r) = \pi(r+1) \\
&C(r,\,s) = s [B + 0.3\cos(6\theta)]\ \label{eq:ex4geom}
\end{split}
\end{align}
\end{minipage}\\[.25cm]
and illustrated in \autoref{fig:overflower}. On the right side of the figure, the boundary conditions and material parameters are defined. The middle surface of the shell features varying principal curvatures and curved boundaries. 
The curved boundaries are clamped and the corresponding conditions (from \autoref{tab:bc}) have to be properly enforced. An analytical solution or reference displacement is not available. Therefore, the error is measured in the strong form of the equilibrium from \autoref{eq:sf} and may be called residual error. In particular, the residual error is the summed element-wise relative $L_2$-error
\begin{align}	\label{eq:reserr}
\varepsilon_{\t{rel,residual}} &= \sum_{i = 1}^{n_\t{Elem}} \varepsilon_{L_2,\t{rel},\tau_i}\\
\varepsilon_{L_2,\t{rel},\tau}^2 &= \dfrac{\int_{\Gamma} \left\lbrace\divG{\tilde{\mat{n}}_\Gamma} + \nG\divG{(\mat{P}\cdot\divG{\mat{m}_\Gamma})} + 2\mat{H}\cdot\divG{\mat{m}_\Gamma} + [\p x_i^\Gamma\mat{H}]_{jk}[\mat{m}_\Gamma]_{ki} +\vek{f}\right\rbrace^2\ \d\Gamma}{\int_{\Gamma} \vek{f}^2\ \d\Gamma} \nonumber\ .
\end{align}
The computation of the residual error requires the evaluation of fourth-order surface derivatives. It is noteworthy that the implementation of these higher-order derivatives is not without efforts. For example, recall that mixed directional surface derivatives are not symmetric. That is, there are $3^4=81$ partial fourth-order derivatives. Nevertheless, if the displacement field is smooth enough this error measure is a suitable quantity for the convergence analysis. \par

\begin{figure}[h]
	\centering
	\subfloat[displacement $\vek{u}$]{\includegraphics[width=.52\textwidth, height = .52\textwidth, keepaspectratio]{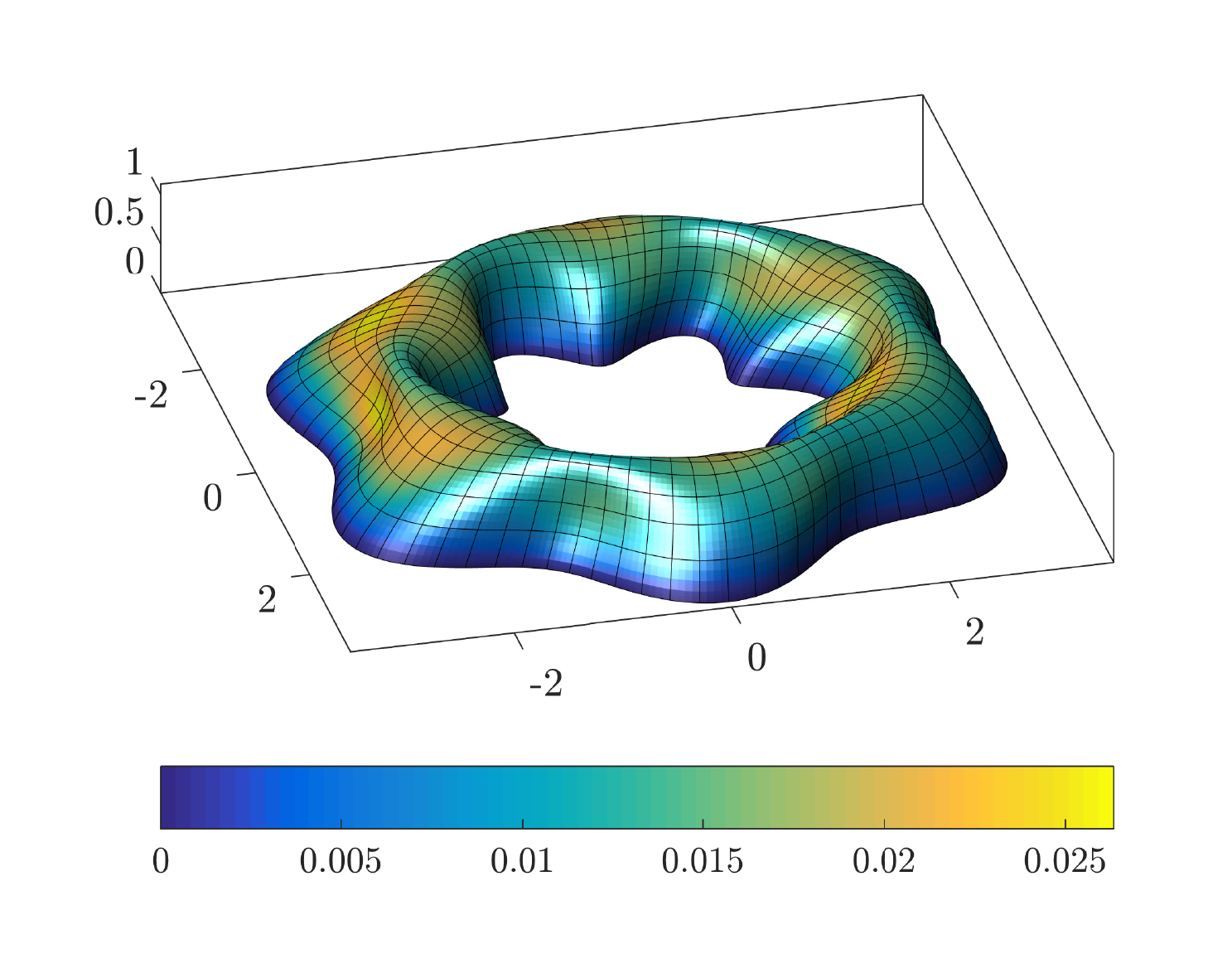}\label{fig:flowershelldisp}}
	\hfil
	\subfloat[convergence]{\includegraphics[width=.45\textwidth, height = .45\textwidth, keepaspectratio]{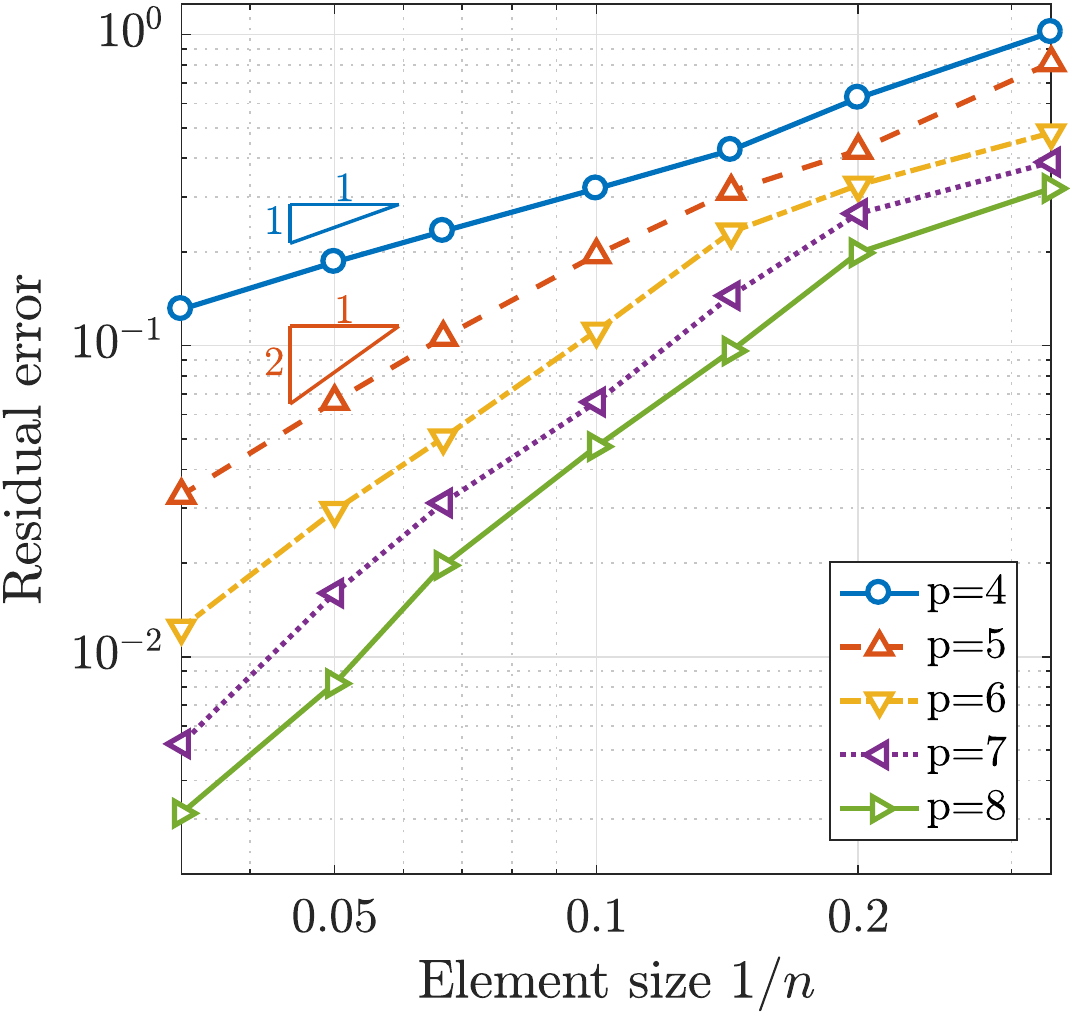}\label{fig:flowershelll2}}
	\caption{Flower shaped shell: (a) Displacement $\vek{u}$ of flower shaped shell (scaled by one order of magnitude), (b) Residual error $\varepsilon_{\t{rel,residual}}$.}
\end{figure}

In \autoref{fig:flowershelldisp}, the deformed shell is illustrated. The displacement field is scaled by one order of magnitude. In \autoref{fig:flowershelll2}, the results of the convergence analysis are plotted. Due to the fact that fourth-order derivatives need to be computed, at least fourth-order shape functions are required. The theoretical optimal order of convergence is $\mathcal{O}(p-3)$ if the solution is smooth enough. One may observe that higher-order convergence rates are achieved, however, rounding-off errors and the conditioning may slightly influence the convergence. Nevertheless, the results are excellent also given the fact that higher-order accurate results for shells (given in double-logarithmic error plots) are the exception.\par

The stored elastic energy at the finest level with a polynomial order $p=8$, which may be seen as an overkill solution, is $\mathfrak{e} = 1.7635958 \pm  \SI{1e-7}{\kilo\newton\metre}$. This stored elastic energy may be used for future benchmark tests, without the need to implement fourth-order derivatives on manifolds.

%% file: Content/Conclusion.tex
\section{Conclusions and Outlook}
\label{sec:conc}

\revStart
The linear Kirchhoff-Love shell theory is reformulated in terms of the TDC using a global Cartesian coordinate system and tensor notation. The resulting model equations apply to shell geometries which are parametrized or not. For example, a parametrization may not be available when shell geometries are implied by the level-set method. Because the TDC-based formulation holds in both cases, it may be seen as a generalization to the classical shell theory which is based on parametrizations and curvilinear coordinates.\par

The TDC-based strong form is used as the starting point to consistently obtain the weak form including all boundary terms well-known in the Kirchhoff-Love theory. Mechanical stress-resultants such as moments, normal and shear forces are defined in global coordinates. Furthermore, the strong form may be used in the numerical results to compute residual errors and thus enable convergence analyses even without the knowledge of exact solutions which, for shells, are scarce.\par

For the discretization, the Surface FEM is used with NURBS as trial and test functions. That is, an isogeometric approach is demonstrated due to continuity requirements. In this case, the presence of a surface mesh (i.e., a NURBS patch), implies a parametrization and although the involved equations and the resulting implementations vary significantly, it is seen that the classical, parametrization-based and the proposed TDC-based formulation are equivalent. For a generic finite element framework enabling various implementations for PDEs on manifolds (in addition to only shells), the TDC-based approach is benefitial, because surface gradients of shape functions may be computed beforehand and are independent of the application.\par

The numerical results confirm higher-order convergence rates. As mentioned, based on the residual errors, a framework for the verification of complex test cases is presented. There is a large potential in the parametrization-free reformulation of shell models, because the obtained PDEs may be discretized with new finite element techniques such as TraceFEM or CutFEM based on implicitly defined surfaces. In this case, neither the problem statement nor the discretization is based on a parametrization.
\revEnd

%% file: Content/AppendixI.tex
\section{Element stiffness matrix}
\label{sec:elemmat}

In order to clarify the implementation, we give the Matlab\textregistered-code of the routine which evaluates the element contribution to the matrix and right hand side.\par

\revStart Using TDC, \revEnd the input contains the shape function data and normal vectors evaluated at the integration points plus the material parameters, see \autoref{lst:ElemContTDC}. Note that the first- and second-order surface derivatives are included in the shape function data, i.e., $\gradG{N}_i(\vek x_j)$ and $\mat{H}\mat{e}^\t{cov}(N_i(\vek{x}_j))$ where $i=1,\,\ldots,\,n$ refers to the $n$ shape functions in the current element (knot span) and $\vek{x}_j$ with $j=1,\,\ldots,\,m$  to integration points. The computation of these quantities is part of the standard finite element technology provided by the implementation and is independent of the application to shells. \par

\revStart
When curvilinear coordinates are used, the input contains shape function data, integration points, material parameters plus the coordinates of the control points of the corresponding element, see \autoref{lst:ElemContClassic}. Note that the included derivatives of the shape functions are the derivatives w.r.t.~reference coordinates, i.e., $\gradR{N}(\vek{r})$. As a first step, the covariant base vectors and the metric tensor in co- and contra-variant form need to be calculated at each integration point and in the next step, the element stiffness matrix and the element load vector are computed.
\revEnd

\includepdf[pages={1-2}, clip,trim=2cm 4.5cm 6cm 2.4cm, nup=2x1,delta= 0.2cm 0cm, width=.6\textwidth, pagecommand={\null\vfill\captionof{lstlisting}{Element contribution in terms of TDC}\label{lst:ElemContTDC}}]{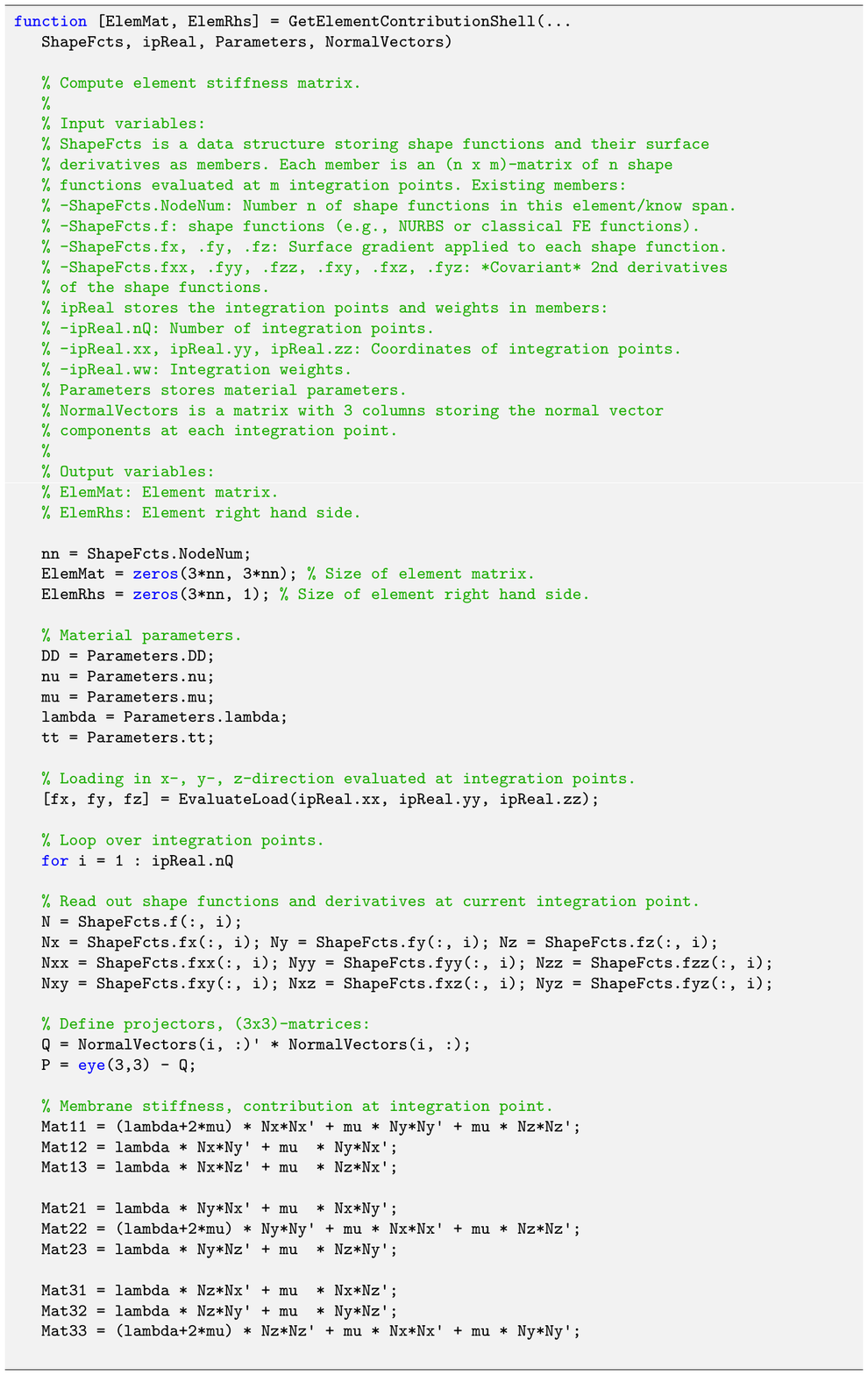}

\includepdf[pages={1-2}, clip,trim=2cm 2.5cm 4.9cm 2.4cm, nup=2x1,delta= 0.2cm 0cm, width=.6\textwidth, pagecommand={\null\vfill\captionof{lstlisting}{Element contribution in terms of local coordinates}\label{lst:ElemContClassic}}]{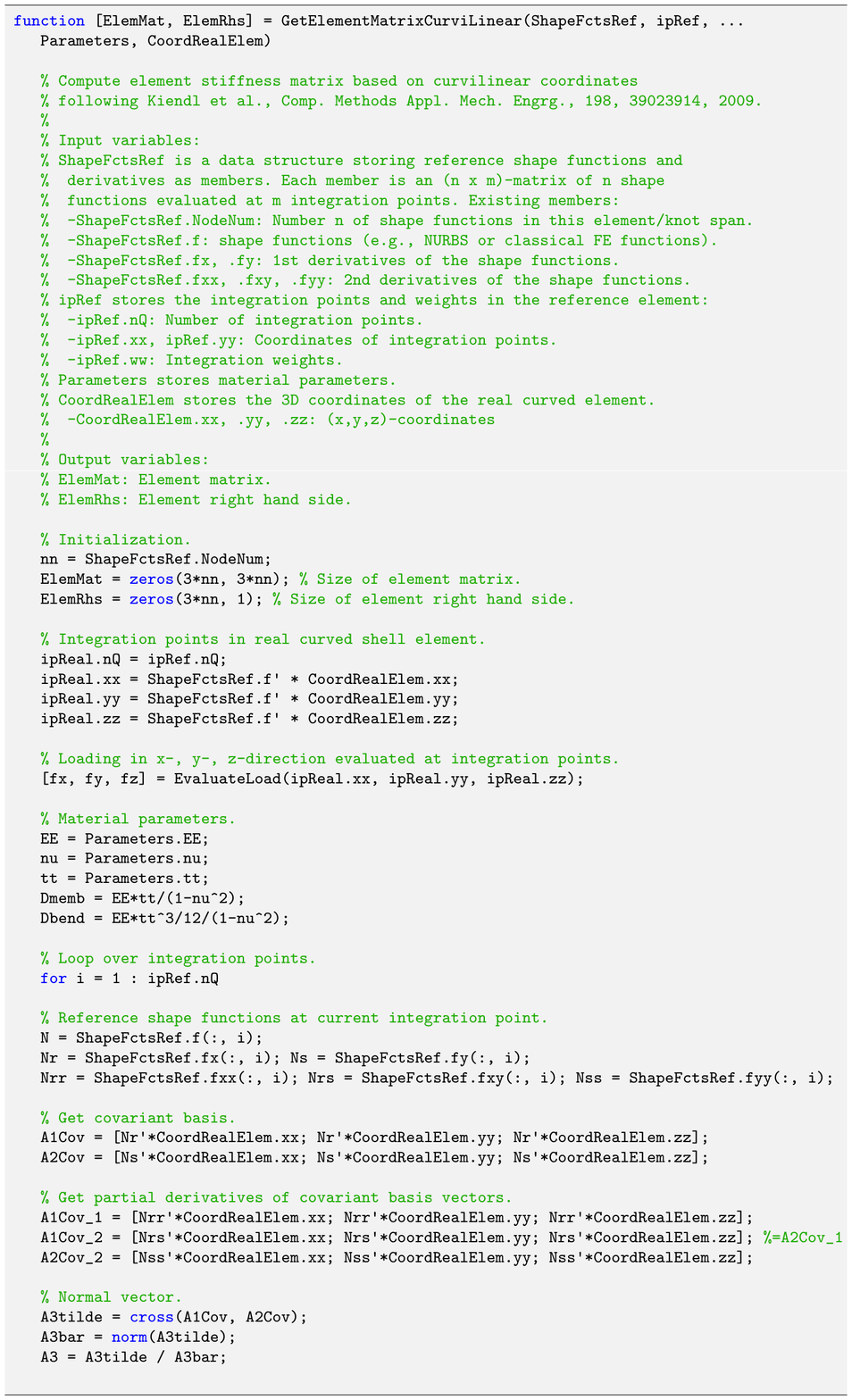}